\RequirePackage{fix-cm}
\documentclass[smallextended]{svjour3}       
\smartqed  
\usepackage{graphicx}
\usepackage{natbib}
\usepackage{amsmath}
\usepackage{amssymb}
\begin{document}

\title{Toward an internally consistent astronomical distance scale}


\author{Richard de Grijs, Fr\'ed\'eric Courbin, Clara
  E. Mart\'inez-V\'azquez, Matteo Monelli, Masamune Oguri, and Sherry
  H. Suyu}

\authorrunning{Richard de Grijs et al.} 

\institute{Richard de Grijs \at Kavli Institute for Astronomy \&
  Astrophysics and Department of Astronomy, Peking University, Yi He
  Yuan Lu 5, Hai Dian District, Beijing 100871, China\\
  \email{grijs@pku.edu.cn};\\
  International Space Science Institute--Beijing, 1 Nanertiao,
  Zhongguancun, Hai Dian District, Beijing 100190, China
\and Fr\'ed\'eric~Courbin \at Institute of Physics, Laboratoire
d'Astrophysique, Ecole Polytechnique F{\'e}d{\'e}rale de Lausanne
(EPFL), Observatoire de Sauverny, CH-1290 Versoix,
Switzerland\\ \email{frederic.courbin@epfl.ch}
\and Clara E. Mart\'inez-V\'azquez \at Instituto de Astrof\'isica de
Canarias, Calle V\'ia L\'actea sn, 38205 La Laguna, Spain
\\ Universidad de La Laguna (ULL), Dpto. Astrof{\'i}sica, E-38206 La
Laguna, Tenerife, Spain\\ \email{clara.marvaz@gmail.com}
\and Matteo Monelli \at Instituto de Astrof\'isica de Canarias, Calle V\'ia
L\'actea sn, 38205 La Laguna, Spain; \\ Universidad de La Laguna
(ULL), Dpto. Astrof{\'i}sica, E-38206 La Laguna, Tenerife, Spain
\\ \email{monelli@iac.es}
\and Masamune Oguri \at Department of Physics, Graduate School of Science,
University of Tokyo, 7-3-1 Hongo, Bunkyo-ku, Tokyo 113-0033, Japan\\
\email{masamune.oguri@ipmu.jp}
\and Sherry~H.~Suyu \at Max-Planck-Institut f{\"u}r Astrophysik,
Karl-Schwarzschild-Str.~1, 85748 Garching, Germany \\
\email{suyu@mpa-garching.mpg.de};\\
Institute of Astronomy and Astrophysics, Academia Sinica, P.O.~Box
23-141, Taipei 10617, Taiwan;\\
Physik-Department, Technische Universit\"at M\"unchen,
James-Franck-Stra\ss{}e~1, 85748 Garching, Germany}

\date{Received: date / Accepted: date}

\maketitle

\begin{abstract}
Accurate astronomical distance determination is crucial for all fields
in astrophysics, from Galactic to cosmological scales. Despite, or
perhaps because of, significant efforts to determine accurate
distances, using a wide range of methods, tracers, and techniques, an
internally consistent astronomical distance framework has not yet been
established. We review current efforts to homogenize the Local Group's
distance framework, with particular emphasis on the potential of RR
Lyrae stars as distance indicators, and attempt to extend this in an
internally consistent manner to cosmological distances. Calibration
based on Type Ia supernovae and distance determinations based on
gravitational lensing represent particularly promising approaches. We
provide a positive outlook to improvements to the status quo expected
from future surveys, missions, and facilities. Astronomical distance
determination has clearly reached maturity and near-consistency.
\keywords{gravitational lensing: strong -- stars: distances -- stars:
  variables: RR Lyrae --- Galaxy: center -- Galaxy: fundamental
  parameters -- galaxies: distances and redshifts -- Local Group --
  Magellanic Clouds -- distance scale}
\end{abstract}

\section{Introduction}\label{sec:intro}

Distance determination from the nearest stars to the edge of the
observable Universe crucially depends on accurate calibration of every
successive rung of the so-called astronomical `distance ladder'
\citep[for a modern version of the distance ladder,
  see][]{degr13}. Local benchmark objects often used to optimally
constrain the uncertainties inherent to using a stepwise calibration
approach include the distances to the Galactic Center and a number of
well-studied galaxies in the Local Group.

$R_0$, the distance from the solar circle to the Galactic Center is
particularly instrumental to achieve a basic calibration for a wide
range of methods used for distance determination, across an extensive
range of distance scales. Among other key physical parameters, the
distances, masses, and luminosities of Galactic objects, as well as
the Milky Way's total mass and luminosity, depend directly on knowing
$R_0$ accurately. For instance, most luminosity and a large number of
mass estimators scale as $R_0^2$, whereas masses based on total
densities or orbital modeling scale as $R_0^3$.

This intrinsic dependence therefore often requires adoption of an
integrated mass and/or a rotation model for the Milky Way, which in
turn would also require us to accurately know the Sun's circular
velocity, $\Theta_0$. As $R_0$ estimates are refined, so are the
estimated distances, masses, and luminosities of numerous Galactic and
extragalactic objects, as well as our best estimates of the rate of
Galactic rotation and the size of the Milky Way. In addition, a highly
accurate direct Galactic Center distance determination is paramount
for a reliable recalibration of the zero points of numerous commonly
used distance calibrators, including of Cepheids, RR Lyrae (see
below), and Mira variable stars.

Beyond Galactic distance tracers, distance measurements to the Large
Magellanic Cloud (LMC) have played an important role in constraining
the value of the Hubble constant, $H_0$. The `{\sl Hubble Space
  Telescope (HST)} Key Project (HSTKP) on the Extragalactic Distance
Scale' \citep{free01} resulted in an $H_0$ estimate of $H_0 = 72 \pm
3$ (statistical) $\pm 7$ (systematic) km s$^{-1}$ Mpc$^{-1}$. The
latter, systematic uncertainty was thought to be dominated by the
remaining systematic uncertainties in the assumed distance to the LMC
\citep{free01,scha08,pietrzynski13}, which contributed of order
$\pm$3--4 km s$^{-1}$ Mpc$^{-1}$ to the tally.

On a number of occasions, the accuracy of LMC distance determinations
has been called into question by claims of `publication bias'
\citep[e.g.,][]{scha08,scha13,rube12,walk12}. Therefore, in
\citet{degr14} we re-analyzed the full body of LMC distance
measurements published between 1990 and 2013, embarking on the most
extensive data-mining effort of this type done to date. Perhaps
somewhat surprisingly, we concluded that strong publication bias is
unlikely to have been the main driver of the clustering of many
published LMC distance moduli. However, we found that many of the
published values were based on highly non-independent tracer samples
and analysis methods. In turn, this interdependence appears to have
led to significant correlations among the body of LMC distances we
considered in \citet{degr14}. Our final, recommended true distance
modulus for the LMC is $(m-M)_0 = 18.49 \pm 0.09$ mag \citep[][see
  also \citealt{cran15}]{degr14}. In an effort to provide a firm mean
distance estimate to the Small Magellanic Cloud (SMC), and thus place
it within the internally consistent Local Group distance framework
(that is, a distance framework where all combinations of distances
lead to the same underlying scaling), \citet{degr15} similarly
performed extensive analysis of the published literature to compile
the largest database available to date containing SMC distance
estimates. For the first time, we provided estimates of the mean SMC
distance based on large numbers of distance tracers, without imposing
any a priori preferences. We derived a true distance modulus of
$(m-M)_0^{\rm SMC} = 18.96 \pm 0.02$, corresponding to a distance of
$61.9^{+0.6}_{-0.5}$ kpc.

Indeed, the nearest galaxies in the Local Group contain numerous
objects that can be used to determine robust distances, that is,
distances with well-defined uncertainties that are mutually consistent
among the different tracers. In \citet{degr14}, \citet{gb14},
\citet{degr15}, and \citet{degr16}, we aimed at establishing an
internally consistent local distance framework by reference to the
subset of distances to the Galactic Center, the LMC and SMC, M31, M32,
and M33, as well as a number of well-known dwarf galaxies. We aimed at
reaching consensus on the best, most homogeneous, and internally most
consistent set of Local Group distance moduli.

\begin{figure}
\begin{center}
\includegraphics[width=\columnwidth]{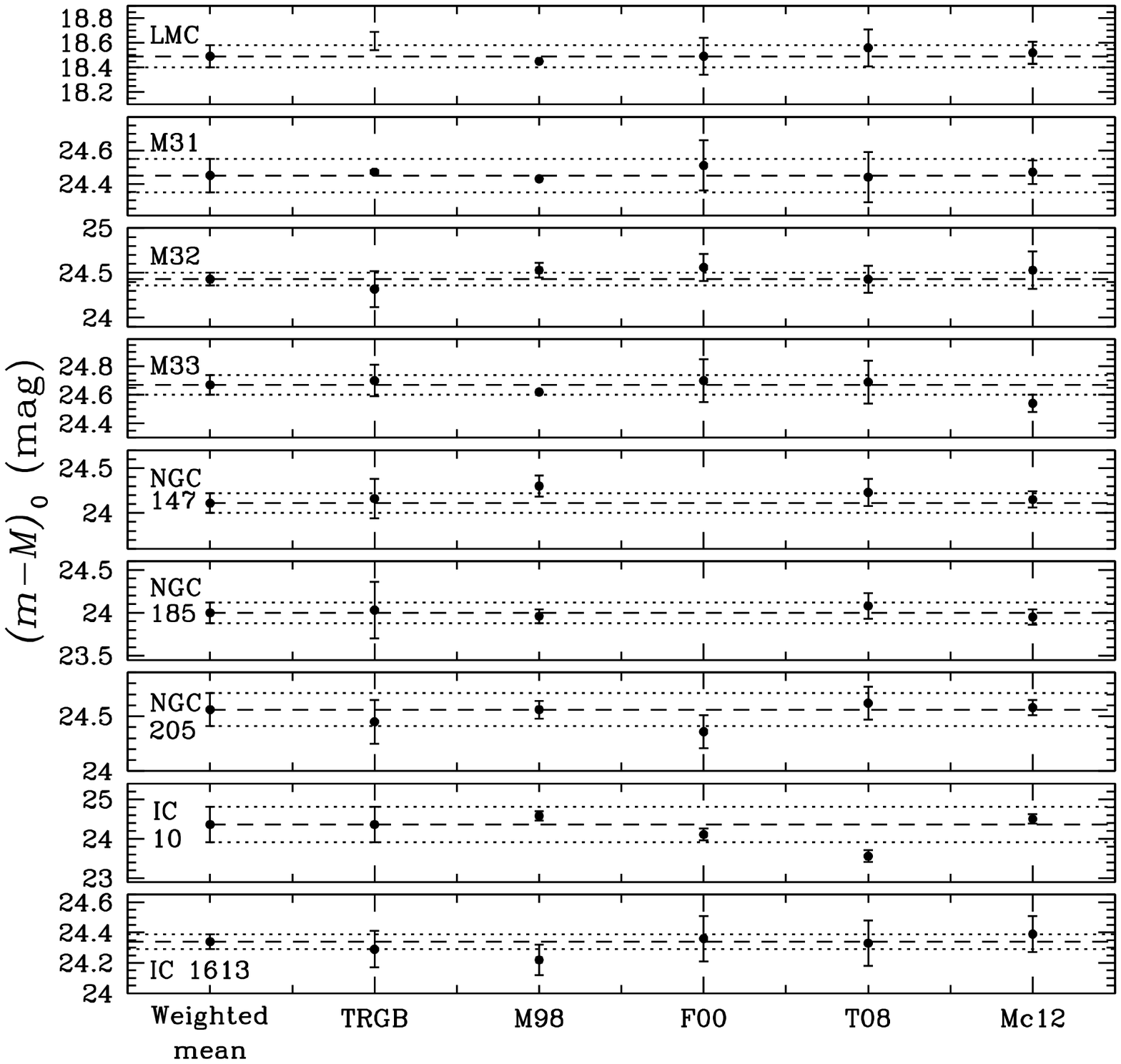}
\caption{Comparison of our set of benchmark distances to the sample of
  Local Group galaxies (indicated in the individual panels) discussed
  in this article with those from a number of recent distance
  compilations. `Weighted mean' and TRGB: \citet{degr15}, except where
  indicated in Table \ref{recommendations.tab}; M98: \citet{mateo98};
  F00: \citet{ferr00}; T08: \citet{tamm08}; Mc12: \citet{mcco12}. From
  \citet{degr15}.}
\label{litcf.fig}
\end{center}
\end{figure}

At the present time, we are in a good position to make recommendations
for the use of robust distance measurements to a set of key Local
Group galaxies: see Table \ref{recommendations.tab} and
Fig. \ref{litcf.fig}.

\begin{table}
\caption{Internally consistent distance moduli (as a function of
  increasing distance) to selected Local Group galaxies, comprising a
  robust local framework \cite[adapted from][]{degr15}.}
\label{recommendations.tab}
\begin{center}
\begin{tabular}{@{}lclcc@{}}
\hline \hline
Galaxy   & $(m-M)_0^{\rm best}$ & Tracer(s) & $(m-M)_0^{\rm TRGB}$ \\
         & (mag)                &           & (mag) \\
\hline
LMC      & $18.49 \pm 0.09$ & Cepheids, RR Lyrae, CMD  & 18.54--18.69     \\
SMC      & $18.96 \pm 0.02$ & EBs, Cepheids, RR Lyrae, TRGB, RC & 
                                                         $19.00 \pm 0.04$ \\
NGC 185  & $24.00 \pm 0.12$ & TRGB, RR Lyrae           & $24.03 \pm 0.33$ \\
NGC 147  & $24.11 \pm 0.11$ & TRGB, RR Lyrae           & $24.16 \pm 0.22$ \\
IC 1613  & $24.34 \pm 0.05$ & Cepheids, RR Lyrae, TRGB & $24.29 \pm 0.12$ \\
IC 10    & $24.36 \pm 0.45$ & TRGB                     & $24.36 \pm 0.45$ \\
M32      & $24.43 \pm 0.07$ & SBF, TRGB, RR Lyrae      & $24.32 \pm 0.20$ \\
M31      & $24.45 \pm 0.10$ & Cepheids, RR Lyrae, TRGB & $24.47 \pm 0.01$ \\
NGC 205  & $24.56 \pm 0.15$ & TRGB, RR Lyrae           & $24.45 \pm 0.20$ \\
M33      & $24.67 \pm 0.07$ & Cepheids, RR Lyrae, TRGB & $24.70 \pm 0.11$ \\
NGC 4258 & $29.29 \pm 0.08$ & H$_2$O masers            & 29.24--29.44     \\
\hline \hline
\end{tabular}
{\flushleft Notes: CMD: color--magnitude diagram; EBs: eclipsing
  binaries; RC: red clump; SBF: surface-brightness fluctuations; TRGB:
  tip of the red giant branch.}
\end{center}
\end{table}

In this review, we will first consider the distances to the Galactic
Center as well as to a select number of Local Group galaxies resulting
from large-scale data mining of the literature (Sections 1--3). Our
aim is to establish an internally consistent distance framework in the
Local Group using multiple, independent methods. We will then proceed
to review the distances to Local Group galaxies based specifically on
space-based data of variable stars, with emphasis on the uniformity in
distance measurement across the nearby galaxies afforded by RR Lyrae
stars (Sections 4--6). Local Group RR Lyrae represent a powerful tool,
because they are found in old stellar populations that are present in
galaxies of any Hubble type and luminosity class (with the possible
exception of the ultra-faint dwarf galaxies). We will discuss a large
number of case studies aimed at their internally consistent
calibration as a single, homogenized distance indicator well beyond
the Local Group.  At greater distances, we focus on efforts to obtain
improved distances and particularly on a number of alternate
techniques to constraining the Hubble constant (Section 7), with
particular emphasis on gravitational lensing as a promising tool
(Section 8). The resulting distances not cross-calibrations of other
distance indicators; instead, they encompass a derived quantity---the
Hubble law---from which distances are computed.

We emphasize that an internally consistent distance scale exists, for
a given cosmology, naturally in the Hubble flow owing to the
relationship between distance and redshift. On these cosmological
scales, the key to consistency is therefore not distance as such, but
re-enforcing the cosmological parameters from which these distances
are derived. At the present time, these cosmological parameters are
predominantly derived on the basis of observations of Type Ia
supernovae (SNe Ia) and the cosmic microwave background. Our
discussion in Sections 7 and 8 focuses on alternative means of
measuring the Hubble constant, not necessarily on the objects'
specific distances. Therefore, gravitational lensing and galaxy
cluster-based techniques are not direct cross-checks on the SNe
Ia-derived distances, but on the Hubble constant measured from SNe
Ia. The latter distance framework is currently calibrated on the basis
of Cepheid distances, which are directly tied into the internally
consistent distance scales discussed in Sections 1--3.

\section{Distance to the Galactic Center}

In recent years various authors have speculated that some degree of
publication bias may have affected Galactic Center distance
determinations
\citep[e.g.,][]{reid89,reid93,niki04,fost10,malk13a,malk13b,fran14}. Aiming
at deriving a statistically well-justified Galactic Center distance
based on a large variety of tracers and reducing any occurrence of
publication bias, \citet{degr16} therefore undertook an extensive
data-mining effort of the literature.

They separated the compilation of $R_0$ determinations into direct and
indirect distance measurements. The former include distances such as
those based on orbital modeling of the `S stars' orbiting Sgr A*, the
closest visual counterpart of the Milky Way's central supermassive
black hole, as well as those relying on statistical parallaxes of
Galactic Center tracers. Careful assessment of the body of published
$R_0$ estimates based on these methods resulted in a Galactic Center
distance recommendation of $R_0 = 8.3 \pm 0.2 \mbox{ (stat.)}  \pm 0.4
\mbox{ (syst.)}$ kpc.

A much larger body of Galactic Center distance determinations is based
on indirect methods, either relying on determinations of the centroids
defined by the three-dimensional distributions of a range of different
tracer populations (e.g., globular clusters, Cepheid, RR Lyrae, or
Mira variables, or red clump stars) or measurements based on kinematic
observations of objects at the solar circle. The latter approaches are
affected by significantly larger uncertainties than the former; the
central, mean Galactic Center distances based on the kinematic methods
are systematically smaller than those based on centroid
determinations. Most centroid-based distances are in good agreement
with those resulting from the direct methods. \citet{degr16} did not
find any conclusive evidence of the presence of publication bias in
the post-1990 $R_0$ measurements.

\section{Internally consistent distances to the Magellanic Clouds}

As the nearest, irregular galaxies to our Milky Way, the Magellanic
Clouds represent the first, important rung of the extragalactic
distance ladder. They cover sizeable areas on the sky, thus providing
access to large samples of tracer objects. For instance, the VISTA
Survey of the Magellanic Clouds \citet{cioni11}, which covers the LMC,
the SMC, the Magellanic Bridge, and a few fields in the Magellanic
Stream, is in currently the process of compiling a 184 deg$^2$
mosaicked map of the entire system.

Indeed, the LMC is the nearest extragalactic environment that
hosts statistically significant samples of the tracer populations for
distance determination, such as Cepheid and RR Lyrae variable stars,
eclipsing binaries (EBs), and red-giant-branch (RGB) stars, as well as
supernova (SN) 1987A. These could thus potentially link the local
(i.e., solar-neighborhood and Galactic) tracers to their counterparts
in more distant and more poorly resolved galaxies. At a distance of
approximately 50 kpc, the LMC represents the only well-studied
environment linking Galactic distance tracers to those in other large
spiral and elliptical galaxies at greater distances. Its
three-dimensional, thin disk-like morphology renders it an ideal
target galaxy where off-center distances can easily be projected onto
the galaxy's center through simple projection.

Lingering systematic uncertainties remain in the distance to the
LMC. This has obvious ramifications in the context of using the LMC
distance as a calibrator, specifically to reduce the uncertainties in
the Hubble constant
\citep[cf.][]{free01,scha08,pietrzynski13}. Although this has also led
to persistent claims of `publication bias' affecting published
distances to the galaxy \citep[cf.][]{scha08,scha13,rube12,walk12},
\citet{degr14} made a case of highly correlated results rather than
publication bias. Nevertheless, the LMC distance moduli is common use
today are likely still affected by poorly understood systematic
uncertainties.

Fortunately, the LMC is sufficiently close that some geometric
distance tracers are readily available, including an increasing sample
of EB systems. Highly promising, \citet{pietrzynski13} determined the
direct distances to eight late-type EB systems in the LMC, resulting
in an average distance modulus of $(m-M)_0= 18.493 \mbox{ stat.} \pm
0.008 \pm 0.047$ (syst.) mag. This value is accurate to 2.2\% and
confirms that the statistically justified distance modulus of
$(m-M)_0= 18.49 \pm 0.09$ is indeed reasonable. \citet{pietrzynski13}
discovered giant stars in EB systems, which uniquely allowed them to
use the very well calibrated surface-brightness--$(V-K)$ color
relation for such stars to determine their angular sizes, so that
their error estimates are more robust and reproducible, and less
affected by lingering systematic effects than competing results based
on early-type EBs. Some of the most recent analyses, e.g.,
\citet{inno16}'s study of the three-dimensional distribution of
classical Cepheids in the LMC's disk, have derived true distance
moduli that are in excellent agreement with both the statistical
average and the geometric distance determination of
\citet{pietrzynski13}. 

In this context it is important to note that \citet{riess16} recently
used Cepheid distances to independently obtain an internally
consistent distance framework including the Milky Way, the LMC, M31,
and the maser host galaxy NGC 4258. This underscores the importance of
the LMC as the lowest rung of the extragalactic distance ladder, and
this also links the local distance framework naturally to our
discussion of derivations of the Hubble constant in Section 7 and
beyond, particularly those based on SNe Ia observations.

\citet{degr15} performed a similar analysis for the SMC as they did
for the LMC. Obtaining a clear-cut distance to the SMC is not as
straightforward as for the LMC, however. First, given the SMC's
bar-like main body with hints of spiral arms and a very extended
`Wing' to the East, defining the position of the galaxy's center is
troublesome. \citet{rube15} recently derived distances to different
areas across the galaxy. They report distances projected onto both the
SMC's kinematic and stellar density centers, $(m-M)_0^{\rm kin} =
18.97 \pm 0.01$ mag and $(m-M)_0^{\rm stars} = 18.91 \pm 0.02$ mag,
respectively. This implies that the choice of SMC center could
introduce systematic uncertainties of order 0.05--0.1 mag in the
resulting distance modulus.

Second, the SMC is very extended along the line of sight. Its depth
could range from 6--12 kpc \citep{crow01} up to 20 kpc \citep[][for
  recent discussions, see e.g.,
  \citealt{kapa12,subr12,cign13,kali13,nide13,scow16}]{groe00},
although the Cepheid population associated with the main body implies
a shallower depth of $1.76 \pm 0.6$ kpc \citep{subr15}. \citep{scow16}
recently also used Cepheids to study the SMC's line-of-sight depth,
but based on observations in nine photometric passbands
simultaneously, including four near- to mid-infrared {\sl Spitzer
  Space Telescope} filters. They confirmed that the Cepheid
distribution is inconsistent with a significant depth, although they
pointed out that the southwestern side of the galaxy is up to 20 kpc
more distant than the SMC's northeast.

Third, the SMC's measured inclination appears to depend on the stellar
tracer (and thus the age of the stellar population) used
\citep{cald86,lane86,groe00,hasc12,subr12,rube15}. Careful geometric
corrections of individual objects back to the galaxy's center will
reduce the scatter in the calibration relations, but this is not
always possible. A warped disk (the presence of which is still
debated) will introduce additional systematic uncertainties, of order
0.1 in the distance modulus \citep{degr15}.

\citet[][his Chapter 6.1.1]{degr11} provides a detailed discussion of
the systematic uncertainties associated with the effects of extinction
in the context of distance determinations. These include uncertainties
related to our knowledge of (i) the most likely extinction law, (ii)
the intrinsic photometric properties of one's calibration objects, and
(iii) the dust geometry. The choice of extinction law is particularly
important when comparing similar types of objects drawn from Galactic
and Magellanic Cloud samples, since `the' Galactic extinction law
(which may vary along different lines of sight) differs systematically
from that in the Magellanic Clouds \citep[for recent studies, see
  e.g.,][and references therein]{doba09,bot10}. Nevertheless, the
differences are generally $\lesssim 0.05$ mag at near-infrared and
longer wavelengths and shortward of $\lambda = 0.8 \mu$m.

\section{RR Lyrae variables in the Local Group}

RR Lyrae stars are low-mass ($\sim 0.6$--$0.8 M_\odot$) core
helium-burning stars in their horizontal-branch (HB) evolutionary
phase. They are radially pulsating variable stars with periods ranging
from 0.2 to 1.0 days and $V$-band amplitudes between 0.2 and
$\lesssim$ 2 mag. Their mean absolute $V$-band magnitude is $M_V\sim
+0.6$ mag, they are of moderate brightness ($\approx 40 L_\odot$), and
their effective mean temperatures range from 6000 to 7250 K
\citep{catelan04b}. RR Lyrae stars are found in stellar systems
hosting old old stellar populations \citep[$t > 10$
  Gyr;][]{walker89,smith95,catelan15}; they can currently be observed
out to distances of $\sim$2 Mpc
\citep{dacosta10,yang14}. Additionally, RR Lyrae stars are primary
distance indicators, since they obey well-defined
optical/near-infrared period--luminosity relations \citep[PLRs; see,
  e.g.,][see for more details also R. Beaton et al., in
  prep.]{bono01d,bono03,clementini03,catelan04b,marconi15}.

The Local Group is home two `normal' spiral galaxies, the Milky Way
and the Andromeda galaxy, M31, shepherded by a number of dwarf
galaxies. Thanks to large photometric surveys, the number of known
satellites of both systems has been increasing rapidly in the last few
years \citep[e.g.][see also
  http://www.astro.uvic.ca/$\sim$alan/Nearby\_Dwarf\_Database.html]{willman06,martin15}. From
an observational point of view, the properties of Local Group galaxies
span the most diverse ranges \citep{mateo98,mcconnachie12}. In
particular, observational approaches to characterize the population of
RR Lyrae have to be adapted to the different systems. Faint nearby
galaxies (often called `ultra-faint dwarfs,' with $-8 \lesssim M_V
\lesssim -1.5$ mag) are characterized by extremely low surface
brightnesses. They typically represent concentrations of a few hundred
stars in a relatively small projected volume covering a few square
arcminutes on the sky. These objects typically have low mean
metallicities ([Fe/H] $\sim -2$ dex), exhibit blue HB morphology, and
host very few RR Lyrae stars \citep[e.g.][]{vivas16b}.

On the other hand, more massive `classical' dwarf satellites ($-13
\lesssim M_V \lesssim -9$ mag) are very rich in RR Lyrae stars
\citep{kinemuchi08,stetson14b,coppola15,martinezvazquez15}, but they
cover large areas on the sky, up to many square degrees, and require
wide-field facilities on medium-sized, ground-based telescopes for
follow-up studies. At present day, a complete census and analysis of
the population of variable stars, and of RR Lyrae stars in particular,
is not yet available for any bright Local Group galaxy \citep[with the
  possible exception of Leo~I and the Carina dwarf spheroidal
  galaxy,][]{stetson14b,coppola15}. A case in point in this context is
illustrated by the monitoring of two of the most massive Milky Way
satellites, the Magellanic Clouds, performed by the Optical
Gravitational Lensing Experiment, which collected $\sim$45,000 RR
Lyrae variables in the Magellanic System in their latest release
\citep{soszynski16}.

Another crucial point to bear in mind is that Local Group galaxies
represent an enormous variety of intrinsic properties, e.g., in terms
of their sizes, morphologies, gas content, chemical evolution, and
stellar populations. As a consequence, they host a variety of families
of variable stars, whose properties reflect the evolutionary history
of the host system. As an example, RR Lyrae stars carry direct
information of the chemical properties of the environment in which
they formed during the first stages of a galaxy's lifecycle. An
intrinsic spread in metallicity in the population of RR Lyrae stars in
a dwarf galaxy implies chemical enrichment which not only occurred at
early times, but which was sufficiently rapid to be imprinted in the
stellar population that we observe today as RR Lyrae stars
\citep{bernard08, martinezvazquez15}. Moreover, the number of RR Lyrae
stars in a galaxy depends on the level of star formation at early
times. Leo A ($M_V=-12.1$ mag), which has very little early
star-formation activity \citep{cole07}, hosts a handful of RR Lyrae
stars \citep{bernard13}, while Sculptor ($M_V=-11.1$ mag), is a
predominantly old system \citep{deboer12a} with a population of RR
Lyrae stars that outnumbers that of Leo A by two orders of magnitude,
despite the latter galaxy's lower mass \citep{martinezvazquez16}.

RR Lyrae stars are primary distance indicators, which can be detected
beyond the limits of the Local Group \citep{dacosta10,
  yang14,mcquinn15e}. Different methods can be applied to derive
distances based on the properties of RR Lyrae stars, including (i) the
luminosity--metallicity relation (LMR) in the $V$ band, (ii) PLRs and
period--Wesenheit relations (PWRs), and (iii) the First Overtone Blue
Edge (FOBE) method. Nevertheless, compiling a homogeneous set of
distances to nearby galaxies using the properties of their variable
stellar populations is not an easy task. Below, we summarize the main
aspects which must be taken into account.

\begin{enumerate}

\item {\bf Photometric calibration:} As regards data pertaining to
  nearby satellite galaxies, large numbers of data sets are available
  in the literature, collected over decades by different observers
  with different telescopes, equipped with different filters and
  cameras. This poses a fundamental problem to derive homogeneous
  photometry in a well-established system
  \citep{stetson00}. Photometric measurements based on {\sl Hubble
    Space Telescope} ({\sl HST}) observations are typically provided
  in the natural {\sc vegamag} photometric system, and they may be
  recalibrated to the Johnson--Cousins system a posteriori.  However,
  different conversion relations have been proposed
  \citep{bedin04,sirianni05,bernard09}, many of which are affected by
  significant color dependences.

\item {\bf Data quality and derivation of the pulsation parameters:}
  Deriving good pulsation properties for variable stars requires
  adequate sampling of their light curves. Especially for existing
  {\sl HST} data, this is not always the case: rarely more than 25
  phase points are available. The parameters derived from sparsely
  populated light curves may depend on the approach used for the
  analysis, the fitting algorithm, and/or the template
  adopted. Moreover, long-term and continuous monitoring of variable
  stars has revealed complex behavior (including period doubling, the
  Blazhko effect, period changes, and non-radial modes, among others)
  which is impossible to detect and characterize on the basis of
  sparsely sampled observations.

\item {\bf The effects of metallicity:} It is worth recalling that
  most relations used to derive distances to RR Lyrae stars require
  one to assume the relevant metallicity. The only notable exceptions
  \citep{marconi15,martinezvazquez15} are the PWR in ($V$, $B-V$) and
  ($V$ , $B-I$). \citet[][their Tables 7 and 8]{marconi15} performed
  an elaborate analysis of optical, optical--near-infrared, and
  near-infrared PWRs. They found that the coefficient of the
  metallicity term was less than $\sim$ 0.05 dex for the ($V$, $B-V$)
  and ($V$, $B-I$) PWRs, indicating that the $V$-band metallicity
  effects are counteracted by the convolution of the reddening vector
  and the $(B-V)$ or $(B-I)$ colors. Since high-resolution
  spectroscopy for RR Lyrae stars in external galaxies is generally
  not available, other indicators or methods must be established to
  estimate the chemical composition of RR Lyrae stars. Typically, the
  metallicity distribution of bright red-giant-branch (RGB) stars is
  available from spectroscopic studies, most commonly low-resolution
  calcium triplet spectra. Since different authors may use different
  metallicity scales or calibrations, deriving homogeneous distances
  is hence not a trivial task.

  Moreover, RGB stars cover an age range of $\gtrsim$1.5 Gyr, and
  the mix of stellar populations of different ages and metallicities
  is highly degenerate in color--magnitude space. The metallicity
  distribution derived from RGB stars is a good approximation of a
  galaxy's global population, but it does not properly describe the RR
  Lyrae stars or the old populations ($>$10 Gyr). As a consequence,
  assuming a mean metallicity for the RR Lyrae stars that is too high
  would introduce a systematic error in distance-modulus estimates, at
  the level of $\sim$0.2 mag. Therefore, assumptions have to be made
  regarding the RR Lyrae metallicity distribution, or alternatively
  this distribution must be determined based on other methods
  \citep{martinezvazquez16}.

\item {\bf Calibration:} Despite its extensive use in the literature
  and its simple form, the LMR relation cannot yet boast a solid
  calibration, neither of its zero point nor of its slope. A number of
  different calibrations exist
  \citep{chaboyer99,bono03,clementini03,carretta09}, and using one
  rather than another can introduce a difference on the order of 0.2
  mag in the distance determination.

\item {\bf Reddening:} Correction for interstellar extinction is a
  crucial step required to derive the absolute magnitudes of one's
  target stars---and in turn their distance moduli. Systems with
  significant internal reddening may be not adequately corrected by
  employing any of the widely used reddening maps
  \citep{schlegel98,schlafly11}. Moreover, differential reddening
  introduces significant scatter in the mean magnitudes and colors of
  stars, which may strongly affect the analysis \citep[e.g.,
    see][]{sarajedini06}. To (at least partially) overcome this
  problem, one can use a Wesenheit pseudo-magnitude
  \citep{vandenbergh68}, which is reddening-free by construction
  (under assumption of a reddening law), or move to near-infrared
  wavelengths, where the effect of extinction is strongly reduced.

\end{enumerate}

RR Lyrae-based distance estimates in the literature are, therefore,
not homogeneous for a large number of reasons: use of different
algorithms for the photometry or of different recipes for the
photometric calibration, the existence of different calibrations for
the same method of distance determination, metallicities which may be
defined on a different scale and/or which may not be appropriate for
the sample of RR Lyrae stars, and the assumed reddening, among
others. Revision of existing data in an effort to provide a
homogeneous distance framework is not a trivial task either. We will
next revise the data and the distance estimates available for galaxies
in the M31 system, for isolated galaxies, and even to some extent for
galaxies outside the Local Group.  We will subsequently proceed, using
already available data, to provide a homogeneous RR Lyrae-based
distance framework. In particular, we will rely on data for six M31
satellite galaxies, which we use as test cases to probe the
reliability of different methods, and we will also provide new
distance estimates to 16 dwarf galaxies based on {\sl HST} data.

\section{RR Lyrae in the Local Group and beyond: status}\label{sec:m31}

The search for RR Lyrae stars in the M31 system has been long limited
because of (i) their (relatively) faint apparent magnitudes ($V\sim
25$ mag) and (ii) crowding effects. The first successful attempt to
identify RR Lyrae stars in the M31 halo was achieved by
\citet{pritchet87a}, using Canada--France-Hawaii Telescope data. A few
years later, \citet{saha90a} and \citet{saha90c} detected candidate RR
Lyrae stars in the M31 dwarf elliptical (dE) satellites NGC 185 and
NGC 147, respectively. Nevertheless, it was only thanks to the {\sl
  HST} and its spatial resolution that it was possible to reach well
below the HB and achieve the first solid determination of the
properties of RR Lyrae stars in the M31 field and its satellites
\citet[e.g.,][]{clementini01}.

\subsection{M31 dwarf spheroidals}\label{sec:dsph}

Based on {\sl HST}/Wide Field and Planetary Camera-2 (WFPC2) data, in
a series of papers devoted to the properties of the stellar
populations in M31 dwarfs, discoveries of RR Lyrae stars were reported
in And~I \citep{dacosta96}, And~II \citep{dacosta00}, and And~III
\citep{dacosta02}. The populations of variable stars detected in these
three galaxies were later analyzed in detail by
\citet[][And~II]{pritzl04} and \citet[][And~I and
  And~III]{pritzl05a}. Moreover, And~VI was studied by
\citet{pritzl02b} on the basis of data of comparable quality. The
distances and distance moduli were derived for the four galaxies using
the LMR relation of \citet{lee90}, assuming a mean metallicity derived
from the mean color of the RGB.

Interestingly, following these pioneering studies, the number of known
satellites of M31 has dramatically increased in the last 10 years,
mostly thanks to the PAndAS survey \citet{mcconnachie09}. However, few
surveys specifically include studies of variable stars. {\sl
  HST}/WFCP2 data were used by \citet{yang12} to study two low-mass
M31 satellites, And~XI and And~XIII, where they detected 17 and 9 bona
fide RR Lyrae candidates, respectively. Their distance moduli were
derived using \citet{chaboyer99}'s calibration of the LMR
relation. The RR Lyrae metallicity was derived using the
\citet{alcock00} relation, and the mean of the distribution was used
to derive the distance modulus.

Other recently discovered satellites have been studied using
wide-field, ground-based Large Binocular Telescope data. The results
have been presented in a series of papers which discuss the RR Lyrae
stellar populations in And~XIX \citep{cusano13}, And~XXI
\citep{cusano15}, and And~XXV \citep{cusano15}. In these studies, the
authors adopted a consistent method to derive the distances to the
three galaxies, based on the LMR relation calibrated by
\citet{clementini03}. In these three cases, the adopted metallicity
was based on low-resolution RGB spectroscopy \citep{collins13}, which
is reasonable for the RR Lyrae stars in these galaxies as well, given
the low mass of the three galaxies.

\subsection{RR Lyrae stars in the M31 field}\label{sec:m31field}

The availability of the {\sl HST}/Advanced Camera for Surveys (ACS)
resulted in a new and significant step forward. New {\sl HST}/ACS data
targeted two of the brightest M31 satellites, M33 \citep{sarajedini06}
and M32 \citep{sarajedini12}. The metallicity of individual RR Lyrae
stars was derived using a relation between [Fe/H] and the period
\citep{sarajedini06} and a relation between [Fe/H], the period, and
the amplitude \citep{alcock00}. Once the mean metallicity of the
sample was known, the distance modulus was derived assuming the LMR as
calibrated by \citet{chaboyer99}. A similar approach was used by
\citet{fiorentino10a} and \citet{fiorentino12a} for M32, who used the
relation of \citet{alcock00} to derive the metallicity distribution of
their M32 sample RR Lyrae stars. However, \citet{fiorentino10a} used
two different approaches to derive the distance modulus to the
galaxy. In addition to the LMR relation \citep{cacciari03}, they also
applied the FOBE method. Given the sizeable sample or RRc-type stars
(first-overtone pulsators), they derived an independent distance
modulus, in agreement with the other LMR method.

The most comprehensive analysis of RR Lyrae stars in M31 has been
published by \citet{jeffery11}, who analyzed different ACS fields in
the galaxy's disk and inner halo, as well as a region overlapping with
the Giant Stellar Stream (GSS) that is being accreted onto M31
\citep{ibata01}. \citet{jeffery11} performed a detailed analysis of
the metallicity determination based on different methods, as well as
their effect on the derived distance modulus. In particular, they
compared the period--metallicity relation \citep{sarajedini06}, the
period--amplitude--metallicity relation \citep{alcock00},
\citet{bono07}'s method based on theoretical
period--amplitude--metallicity relations, and the period--metallicity
relation for RRc type stars \citep{sandage93d}. A comparison of the
metallicities derived photometrically using the first three methods
with respect of spectroscopic measurements reveals good agreement
between the \citet{alcock00} and \citet{bono07} methods, although the
latter exhibits a significantly smaller dispersion.

Overall, very little is known about the global properties of the RR
Lyrae stars in M31. The {\sl HST} data available only cover a tiny
fraction of its body, and they are all concentrated in the inner halo
\citep{ferguson16}. Moreover, as many satellite objects (such as NGC
205, the GSS, or And~I) are embedded in, or projected along the line
of sight, it is often difficult to unequivocally distinguish variables
associated with either M31 or its satellite galaxies.

\subsection{Isolated Dwarfs}\label{sec:isol}

At present, few Local Group galaxies are really isolated and located
outside the virial radii of either the Milky Way or M31
\citep{mcconnachie12}. Nevertheless, a handful (DDO~210, VV~124,
Sagittarius dwarf irregular galaxy) are truly isolated systems in the
sense that they are currently on their first approach of the innermost
region of the Local Group, and they never experienced any strong
interactions with either the Milky Way or M31. For some other
galaxies, the situation is not as clear, because they are currently
located outside the tidal radius of either major spiral galaxy, but we
cannot exclude the possibility of past interactions
\citep{fraternali09}. Detailed color--magnitude analysis reaching the
old main-sequence turn-off of relatively and truly isolated Local
Group dwarf galaxies requires use of the {\sl HST}. The first
systematic investigation of a representative sample of such galaxies
has been performed by the LCID (Local Cosmology from Isolated Dwarfs)
team \citep{gallart15}. Recently, more data have been acquired for
DDO~210 \citep{cole14}, WLM, and Pegasus dwarf irregular (dIrr) galaxy
(as yet unpublished). Studies of variable stars have been published by
\citet[][Cetus and Tucana]{bernard09}, \citet[][IC~1613]{bernard10},
\citet[][Leo~A]{bernard13}, \citet[][Leo~T]{clementini12},
\citet[][Phoenix]{ordonez14}, and \citet[][DDO~210]{ordonez16}.

\subsubsection{Purely old systems: Cetus and Tucana}

Cetus and Tucana represent remarkable examples of isolated dwarf
spheroidal (dSph) galaxies, since they are the only two gas-poor,
purely old systems that do not follow the so-called
density--morphology relation. In fact, the general tendency is that
dSph systems are more clustered around large spirals, while dIrr
galaxies are usually found in more isolated environments. However,
Cetus and Tucana are the only two known purely old dwarf galaxies
located at many hundreds of kpc from both the Milky Way and M31
\citep{monelli10b}. Nevertheless, their recessional velocities suggest
that they may have been involved in an interaction a few Gyr ago
\citep{sales07b,donghia09}. \citet{bernard09} analyzed the detailed
properties of RR Lyrae stars in both systems. Distance moduli were
homogeneously derived by means of the LMR relation based on the
\citet{clementini03} calibration and the FOBE method.

Despite many similarities among the general properties of their
stellar populations, their star-formation histories (SFHs) were
different during the oldest epochs. In particular, although the
initial epoch of star formation was the same, the peak of the
star-formation activity occurred at an earlier epoch in Tucana than in
Cetus \citep{monelli10c}. This apparently subtle difference clearly
emerges from the properties of their RR Lyrae stars. On the one hand,
\citet{monelli12b} presented a sample of more than 600 RR Lyrae stars
in Cetus using wide-field Very Large Telescope (VLT)/VIMOS data and
concluded that the properties of the RR Lyrae stars in Cetus are
homogeneous over the entire galaxy body. The mean period, magnitude,
and amplitude of the RR Lyrae stars do not significantly change as a
function of radius, and overall, RR Lyrae stars seem to originate from
a stellar population with similar chemical properties.

On the other hand, Tucana represents a much more complex
case. \citet{bernard08} discussed the complex properties of a few
hundred RR Lyrae stars, showing that the samples of bright and faint
RR Lyrae stars have different pulsation properties and spatial
distributions: faint stars are more centrally concentrated and their
periods are shorter, at fixed amplitude. These results can be
explained naturally if RR Lyrae stars exhibit an intrinsic spread in
their metallicity distribution: faint RR Lyrae stars belong to a
slightly more metal-rich population, consistent with them being
preferentially formed in the innermost region of the galaxy, as
commonly found in most nearby dSph galaxies.

These two examples strongly suggest that the detailed properties of RR
Lyrae stars must be well-known in order to safely use them as distance
indicators. If the population of RR Lyrae stars presents an intrinsic
spread in metallicity (e.g., IC 1613, \citealt{bernard10}; Sculptor,
\citealt{martinezvazquez15}; NGC 185, Monelli et al. submitted), it is
risky to blindly adopt a mean luminosity and a mean metallicity for
the entire sample. \citet{bernard09} derived the distance modulus of
Tucana using the bright and the faint samples of RR Lyrae stars
separately, assuming an appropriate metallicity constraint resulting
from their SFH calculation, and they obtained consistent distance
moduli, within the uncertainties.

\subsubsection{Galaxies with extended SFHs}

\begin{enumerate}

\item {\bf IC 1613 and Leo A:} Whenever the SFH of a galaxy covers the
  full cosmic time, it is possible to use both old RR Lyrae stars and
  young classical Cepheids (CCs) to simultaneously obtain distance
  estimates. Examples of such analyses have been presented for IC 1613
  \citep{bernard10} and Leo A \citep{bernard13}. These authors used
  consistent assumptions and five different relations to derive their
  distance moduli, including period--luminosity and
  period--luminosity--color relations for CCs, combined with the LMR
  and FOBE methods for the RR Lyrae stars. Interestingly, while good
  agreement was found for IC~1613, the distance derived based on CCs
  was somewhat longer than that derived from RR Lyrae stars.

  In addition, a number of issues should be taken into account in
  assessing similar analyses. The periods of Cepheids can span from a
  few to hundreds of days. Therefore, it is very difficult, from an
  observational point of view, to collect a data set which allows for
  a good characterization of the light curves over the full period
  range, especially using {\sl HST}. Nevertheless, coupling extensive
  ground-based data sets with dedicated and optimized space-borne
  observations has provided an excellent basis for such
  investigations. \citet{scowcroft13} provided a multiwavelength view
  of Cepheids in IC 1613, combining optical results from the OGLE
  survey \citep{udalski01} with multi-epoch observations in $JHK_{\rm
    s}$, obtained with the Magellan telescope, and at longer
  wavelengths using the {\sl Spitzer Space Telescope}. The
  well-characterized optical light curves allowed them to detect and
  constrain the infrared properties of a Cepheid sample in IC 1613 and
  derive PLRs. The inclusion of deep {\sl HST}/WFC3 near-infrared data
  allowed \citet{hatt17} to extend the comparison to RR Lyrae
  stars. These latter authors obtained very good agreement between
  Population I and Population II distance indicators. The same
  approach was used by \citet{rich14} to study the Cepheid population
  of NGC 6822. Combining optical OGLE detections with ground-based
  near-infrared and {\sl Spitzer} mid-infrared data, \citet{rich14}
  derived six PLRs, as well as self-consistent distance and reddening
  estimates.

\item {\bf DDO~210 and Phoenix:} DDO~210 is a low-mass, isolated
  galaxy with properties similar to those of Leo A \citep{cole14}. Its
  variable-star content has been studied by \citet{ordonez16}, who
  derived a distance modulus using the period--luminosity--metallicity
  relation in the $I$ band of \citet{catelan04b}, once they had
  derived individual metallicities using the \citet{alcock00}
  relation.

  \citet{gallart04b} presented a large number of candidate RR Lyrae
  stars in Phoenix, but the data were not good enough for a detailed
  analysis. Such an analysis was, however, performed using {\sl
    HST}/WFPC2 images by \citet{ordonez14}, who used RR Lyrae stars to
  study the early chemical evolution of Phoenix. Unfortunately,
  neither paper used RR Lyrae stars to derive distances.

  Cepheids have been identified in both galaxies. However, the
  identification of CCs, especially in the short-period tail of the
  distribution, can be complicated by the presence of anomalous
  Cepheids (ACs). ACs and CCs both cross the instability strip during
  the central helium-burning phase of their evolution. The key
  difference with CCs is that ACs ignite helium in a degenerate core,
  while CCs are sufficiently massive that this occurs under
  non-degenerate conditions. Therefore, ACs have masses $\lesssim 2.3
  M_{\odot}$ and low metallicities ([Fe/H]$\lesssim -1.5$ dex;
  \citealt{fiorentino06}), since in this mass range, stars with higher
  metallicity evolve on the red side of the instability strip. Phoenix
  was the first system where CCs and ACs were found to coexist
  \citep{gallart04b}. The coexistence of the two types of Cepheids has
  since been confirmed in other low-metallicity systems, including IC
  1613 \citep{bernard10}, Leo A \citep{bernard13}, Leo~I
  \citep{fiorentino12d}, and DDO~210 \citep{ordonez16}. Also, both
  Magellanic Clouds contain a small sample of ACs
  \citep{soszynski08c,soszynski16}.

\end{enumerate}

\subsubsection{Leo~T} 

Leo~T is an intriguing low-mass, star-forming, relatively isolated
dwarf galaxy. Using {\sl HST}/WFPC2 data, \citet{clementini12}
detected one RR Lyrae star and 12 ACs. This is because of the
combination of its small total baryonic mass and its SFH, which is
characterized by a dominant episode at intermediate ages and a low
rate at early epochs. Nevertheless, the properties of the only RR
Lyrae star discovered allowed an estimate of the distance modulus,
based on the \citet{clementini03} LMR, adopting a proxy metallicity,
spectroscopically derived for the galaxy's RGB stars.

\subsection{RR Lyrae beyond the Local Group}\label{sec:outside}

The {\sl HST} has allowed the discovery of RR Lyrae stars in a handful
of dwarf galaxies outside the Local Group. \citet{dacosta10} first
reported detections of RR Lyrae in two dwarf galaxies in the Sculptor
group. Variables were later analyzed in detail by
\citet{yang14}. Individual metallicities were derived using the
\citet{alcock00} method, and the mean metallicity of the distribution
was used with the \citet{chaboyer99} LMR relation. The derived
distances, close to 2 Mpc, are in agreement, within the uncertainties,
with other estimates based on the TRGB.

The only other galaxy external to the Local Group where RR Lyrae have
been discovered is Leo P \citep{mcquinn15e}. This is a very
interesting low-mass, gas-rich galaxy with an extended SFH, similar to
Leo~T. A distance modulus based on the 10 RR Lyrae discovered was
obtained using the \citet{carretta00b} LMR relation. The metallicity
pertaining to the RR Lyrae stars was estimated by comparing the
distribution of stars in the color--magnitude diagram with theoretical
isochrones.

\section{Toward homogeneous RR Lyrae-based distance determinations}

\subsection{The ISLAndS project}\label{sec:distances}

ISLAndS is an ongoing project targeting six M31 satellites with as
main aim to compare the SFHs in the Milky Way and the M31 systems
\citep{weisz14a,monelli16,skillman16}. Full analysis of the variable
stars will be presented in a series of forthcoming papers
(Mart\'inez-V\'azquez et al., in prep). Key results regarding distance
estimates are summarized here. The sample of galaxies includes And~I,
And~II, And~III, And~XV, And~XVI, and And~XXVIII, thus spanning a wide
range in mass, luminosity, and distance from M31. Next, we will use
the ISLAndS galaxies as test cases to compare the performance of four
different methods to derive their distance moduli. Three of the
methods are based on the properties of the RR Lyrae stars, including
(i) the reddening-free PWR \citep{marconi15}, (ii) the LMR
\citep{bono03,clementini03}, and (iii) the FOBE relation
\citep{caputo00c}; these are supplemented by (iv) the TRGB method.

As a first step, the metallicity values were homogenized to
\citet{carretta09} scale. For those galaxies with metallicity
estimates based on theoretical spectra, we applied a correction to
take into account the updated solar iron abundance, rescaling to
log$\epsilon_{\rm Fe}$ = 7.54 dex.

The second step is meant to ensure homogeneous assumptions for the
metallicity. The metallicity estimates available in the literature
were derived using calcium-triplet spectroscopy of bright RGB
stars. The mean metallicity of the RGB stars in And~III
\citep{kalirai10}, And~XV \citep{letarte09}, And~XVI
\citep{collins15}, and And~XXVIII \citep{slater15}, is close to
[Fe/H]$\sim -1.8$ dex (or less). In agreement with the limited age and
metallicity spreads characteristic of the dominant population, derived
from the SFHs \citep{skillman16}, we assume that the spectroscopic mean
metallicity is representative of the old population as well.

However, And~I and And~II exhibit higher mean metallicities and larger
metallicity spreads \citep{ho12,ho15}. Nevertheless, the small number
of high-amplitude, short-period (fundamental-mode) RRab-type stars in
these galaxies \citep{fiorentino15a} suggests that, even if the tail
of the RR Lyrae metallicity distribution reaches such relatively high
values, the bulk of the RR Lyrae stars must have a lower metallicity
([Fe/H]$< -1.5$ dex; \citealt{fiorentino15a}). Therefore, we adopted
[Fe/H]= $-1.8$ dex as representative of the mean metallicity of the RR
Lyrae stars in And~I and And~II.

Once the metallicity had been fixed, we estimated the distance moduli
to the ISLAndS galaxies as outlined below. The results are summarized
in Fig.  \ref{fig:dist} and Table \ref{tab:table}.

\begin{enumerate}

\item {\bf Period--Wesenheit relations:} Since the data were
  originally calibrated according to the VEGAMAG system, individual
  phase points were recalibrated to the Johnson system using the
  transformation provided by \citet{bernard09} and optimized for the
  RR Lyrae color range. Since the F475W filter is characterized by a
  wide passband, similar to the $g$ band, measurements were calibrated
  in both the $B$ and and the $V$ filters, while the F814W band was
  transformed to the Johnson--Cousins $I$ filter. This allowed us to
  use two PWRs from \citet{marconi15}, the ($I$, $B-I$) and ($I$,
  $V-I$) relations, but not the metallicity-independent ($V$, $B-V$)
  PWR, because the $B$ and $V$ measurements are not independent. For
  the adopted relations, a metallicity dependence is present, but it
  is weak: a change of 0.3 dex translates into a change in the
  distance modulus of order 0.03 mag.

  We applied both relations to the samples of RRab- and RRc-type
  variables, as well as to the total sample (RRab + fundamentalized
  RRc: $\log P_{\rm fund} = \log P_{\rm RRc} + 0.127$;
  \citealt{bono01d}). The comparison shows very good agreement among
  the different determinations, on average within $\pm$0.04 mag. Since
  the number of stars in the full sample is the largest, we selected
  as final adopted distance modulus the mean value of the distance
  moduli calculated using both PWRs and that for the full sample.
  Table~\ref{tab:table} presents for the six galaxies (Col. 1) the
  range of distance moduli available in the literature (Col. 2), and
  our final estimate based on the PWR of the full sample of RR Lyrae
  stars (Col. 3).

\item {\bf Luminosity--metallicity relation:} We adopted two different
  relations, i.e., those proposed by \citet{clementini03} and
  \citet{bono03}. The zero-point of the former was modified such that
  the distance modulus to the LMC is in agreement with the distance
  modulus obtained by \citet{pietrzynski13}, assuming [Fe/H]= $-1.5$
  dex for the RR Lyrae stars \citep{gratton04b}. The resulting
  equation is therefore
  \begin{equation}\label{eq:c03}
  M_{V,\rm C03}= 0.892 \pm0.052 + (0.214 \pm 0.047) {\rm [Fe/H]}.
  \end{equation}

\begin{figure*}
    \hspace{-0.5cm}
    \includegraphics[scale=0.65]{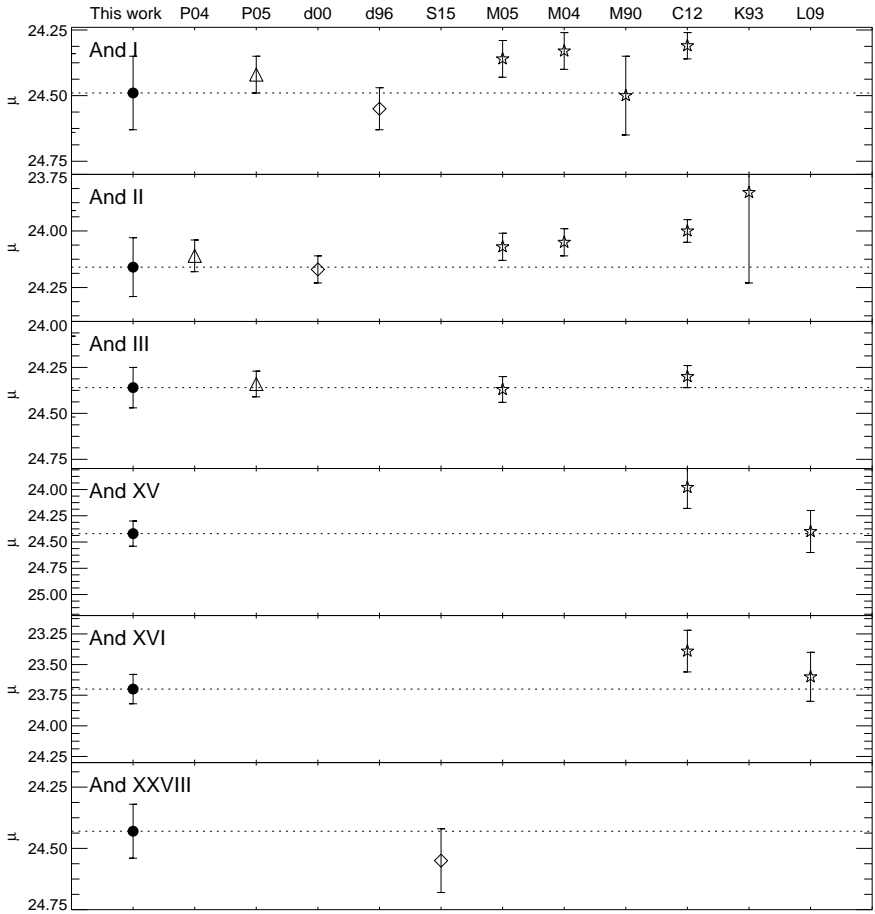}
    \caption{Summary of our distance modulus estimates (solid symbols)
      and comparison with literature values (open symbols). Estimates
      based on the TRGB are provided only for the most massive
      galaxies, for which the TRGB could be estimated reliably. The
      solid circles and dotted lines show the measurements based on
      the PWR relations. Other symbols show values taken from the
      literature, for comparison. In particular, we report values
      based on RR Lyrae stars
      \citep[asterisks:][]{pritzl02b,pritzl05a}, the HB luminosity
      \citep[open squares:][]{dacosta96, dacosta00, slater15}, and the
      TRGB \citep[open diamonds:][]{mould90, koenig93, mcconnachie04a,
        mcconnachie05, conn12}.}
    \label{fig:dist}
\end{figure*}


  The second calibration adopted was a double-linear relation with a
  break in the slope at [Fe/H]= $-1.6$ dex. We adopted the lower
  metallicity, appropriate for the six ISLAndS galaxies ([Fe/H]= $-1.8$
  or less).

\item {\bf The FOBE method:} This method is based on the predicted
  period--luminosity--metallicity relation for pulsators located along
  the FOBE of the instability strip \citep[see][]{caputo00c}:

  \begin{equation}\label{eq:fobe}
  M_{V,\rm FOBE}= -0.685 -2.255 \log(P_{\rm FOBE}) -1.259
  \log(M/M_{\odot}) +0.058 \log(Z).
  \end{equation}

  This is a well-defined technique for stellar systems with a large
  number of RRc-type stars, in particular if the blue side of the
  first-overtone instability strip is well-populated. For this reason,
  we applied it only to five of our six galaxies, since the small
  number of RRc stars in And~XVI may yield inconsistent results
  \citep{monelli16}. To derive the stellar masses, we adopted the
  BaSTI evolutionary models \citep{pietrinferni04}. For the selected
  metallicity, a star with an effective temperature typical of RRc
  stars ($\log T_{\rm eff}\approx 3.86$ [K]), we derived values of $M
  \sim 0.7 M_{\odot}$. Distance moduli obtained for each galaxy using
  this method are included in Fig. \ref{fig:dist}.

\begin{table}
\begin{tabular}{p{2cm}p{2.4cm}p{3cm}}
\hline \hline
\label{tab:table}
Galaxy       & \multicolumn{2}{c}{True distance moduli} \\
\cline{2-3}\\
             & Literature value    &          PWR       \\
             & (mag)               & (mag) \\
\hline
And~I        &  24.31--24.55   &   24.51$\pm$0.08   \\
And~II       &  23.87--24.17   &   24.17$\pm$0.07   \\
And~III      &  24.30--24.37   &   24.38$\pm$0.06   \\
And~XV       &  23.98--24.4    &   24.43$\pm$0.07   \\
And~XVI      &  23.39--23.6    &   23.71$\pm$0.07   \\
And~XXVIII   &  24.55$\pm$0.13     &   24.45$\pm$0.05   \\
\hline \hline
\end{tabular}
\end{table}

\item {\bf Tip of the RGB:} The TRGB is a well-studied, work-horse
  standard candle by virtue of its weak dependence on both age
  \citep{salaris02} and, particularly in the $I$ band, on the
  metallicity \citep[at least for relatively metal-poor
    systems;][]{dacosta90,lee93}. The TRGB is frequently used to
  obtain distance estimates to galaxies of all morphological types
  \citep[e.g.,][]{rizzi07, bellazzini11a, wu14}. However, determining
  the cut-off in the luminosity function at the bright end of the RGB
  is not straightforward in low-mass systems because of the inherently
  small number of bright RGB stars \citep{madore95, bellazzini02,
    bellazzini08}. A total of more than about 100 stars in the top
  magnitude range of the RGB is considered a safe threshold for
  reliable estimates. In our galaxies, this condition is met only by
  And~I ($N>$200), And~II ($N>$150), and nearly in And~III
  ($N\sim$90). The small numbers of such stars in the other three
  galaxies prevent us from deriving reliable measurements of the
  apparent magnitudes of their TRGBs. We adopted the calibrations of
  \citet{rizzi07}, \citet{bellazzini11a} and \citet{cassisi13c} to
  compile Fig. \ref{fig:dist}.

\end{enumerate}

\begin{figure*}
    \hspace{2cm}
    \includegraphics[scale=0.4]{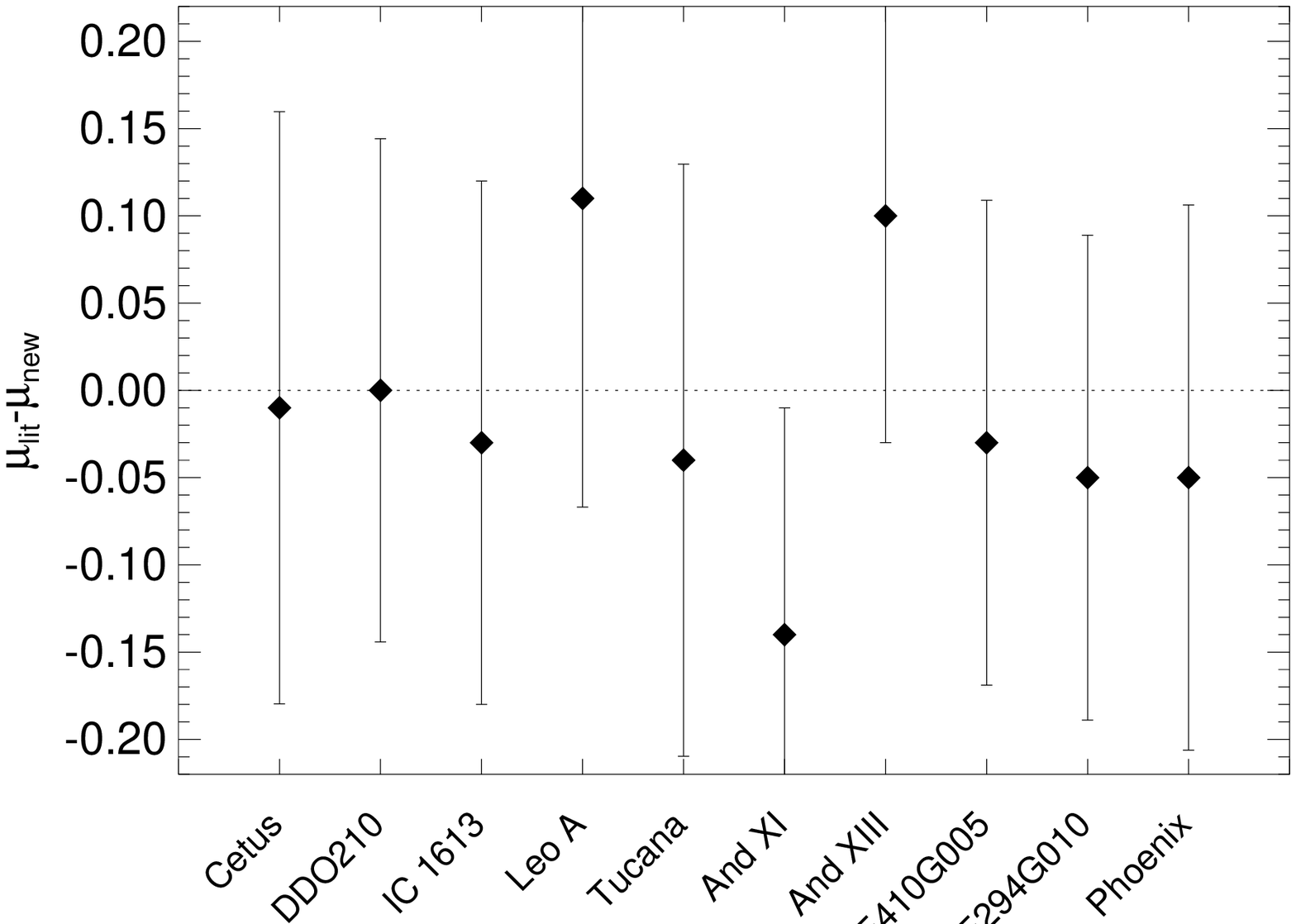}

    \caption{Differences between distance-modulus estimates obtained
      based on RR Lyrae stars and literature values for a number of
      galaxies. These updated values represent homogenized
      metallicities, employing the \citet{marconi15} PWR.}

    \label{fig:distnew}
\end{figure*}


Figure \ref{fig:dist} summarizes the distance modulus determinations
derived in this paper. In particular, the solid circles and the dotted
line show the adopted final distance moduli based on averaging the
values from the two PWRs. Filled diamonds show the results from the
other methods. This figure shows that the agreement among the
different methods employed here is remarkably good, since most of the
derived distance moduli agree within 1$\sigma$. Taking as reference
the PWR distance, some general trends can be discerned. The distance
derived using the LMR based on the \citet{bono03} calibration is
marginally longer with respect to that resulting from the
\citet{clementini03} calibration. The FOBE distance is longer than the
PWR distance in three cases (And~III, And~XV, and And~XVI) and shorter
for And~II.  Nevertheless, this method is most sensitive to the
sampling of the instability strip, and in particular the lack of RR
Lyrae close to the blue edge of the instability strip introduces a
bias toward greater distances.

Asterisks and open symbols in Fig. \ref{fig:dist} represent distance
modulus estimates available in the literature, derived using different
techniques, including those based on RR Lyrae stars
\citep[asterisks:][]{pritzl02a,pritzl05b}, the HB luminosity
\citep[open squares:][]{dacosta96, dacosta00, slater15}, and the TRGB
\citep[open diamonds:][]{mould90, koenig93, mcconnachie04a,
  mcconnachie05, conn12}. Figure \ref{fig:dist} shows generally good
agreement with our estimates, within the uncertainties. Note that the
TRGB tends to yield shorter distances than the RR Lyrae and the HB
luminosity-based methods, although they are still compatible within
1.5$\sigma$. A number of discrepant cases (And~XV and And~XVI; see
\citealt{conn12}) can be ascribed to the sparsely populated bright
part of the RGB in these galaxies.

\subsection{Revising the distances to Local Group galaxies}\label{sec:revision}

Using existing data to construct a homogeneous distance framework for
a large number of Local Group galaxies is still far from having been
achieved. Ideally, one would want to derive homogeneous pulsation
properties and metallicities, and then uniformly use RR Lyrae stars to
derive distances. However, the highly inhomogeneous data set available
(in terms of data quality, filter passbands used, and completeness)
renders this a complicated task. Moreover, different authors often use
different methods of data analysis and, critical in this case,
different approaches and assumptions as regards the metallicity.

\begin{table}
\begin{tabular}{p{2cm}p{2.4cm}p{3cm}p{3cm}}
\hline \hline
\label{tab:table2}
Galaxy       & \multicolumn{3}{c}{Distance moduli}\\
\cline{2-4}\\
             & Literature         &          PWR                 & LMR \\
             & (mag)              & (mag)                        & (mag)\\
\hline
Cetus     &  24.46$\pm$0.12$^a$  &   24.47$\pm$0.02$\pm$0.04  & 24.43$\pm$0.08 \\  
DDO~210      &  25.07$\pm$0.12$^b$  &   25.07$\pm$0.03$\pm$0.05  & --- \\  
IC~1613      &  24.44$\pm$0.09$^c$  &   24.47$\pm$0.02$\pm$0.04  & 23.41$\pm$0.08 \\  
Leo A     &  24.57$\pm$0.13$^d$  &   24.46$\pm$0.07$\pm$0.10  & 24.41$\pm$0.08 \\  
Tucana       &  24.74$\pm$0.12$^e$  &   24.78$\pm$0.01$\pm$0.05  & 24.72$\pm$0.08 \\  
And~XI       &  24.33$\pm$0.05$^f$  &   24.47$\pm$0.05$\pm$0.16  & 24.55$\pm$0.08 \\  
And~XIII     &  24.62$\pm$0.05$^g$  &   24.52$\pm$0.07$\pm$0.14  & 24.72$\pm$0.08 \\  
ESO410-G005  &  26.33$\pm$0.07$^h$  &   26.36$\pm$0.01$\pm$0.23  & 26.36$\pm$0.08 \\  
ESO294-G010  &  26.40$\pm$0.07$^i$  &   26.45$\pm$0.01$\pm$0.20  & 26.42$\pm$0.08 \\  
Phoenix      &  23.09$\pm$0.10$^j$  &   23.14$\pm$0.02$\pm$0.15  & 23.17$\pm$0.08 \\  
\hline \hline
\end{tabular}
\begin{scriptsize}
$^a$\citet{bernard09},  
$^b$\citet{ordonez15},
  $^c$\citet{bernard13},
$^d$\citet{bernard13},  
$^e$\citet{bernard09},  
$^f$\citet{yang12}, 
$^g$\citet{yang12}, 
$^h$\citet{yang14}, 
$^i$\citet{yang14}, 
$^j$\citet{hidalgo09}.  
\end{scriptsize}
\end{table}

In an incomplete attempt to resolve this situation, based on a similar
approach as that adopted for the ISLAndS sample, we re-analyzed the
available data for a sample of Local Group galaxies for which
high-quality {\sl HST} data are available. This includes galaxies
characterized by different levels of isolation and extends to systems
in the Sculptor group. We selected those galaxies for which photometry
in the Johnson $V$ and $I$ or in the $B$ and $I$ bands is available in
the literature.\footnote{We emphasize that we did not attempt to use
  the original photometry in the VEGAMAG system and transform it,
  since the original data are often not provided. Therefore, the
  photometric conversion to the Johnson system may have been performed
  using different methods by different authors. Although we do not
  expect this to introduce significant systematics in the distance
  determination, this issue must be taken into account for future
  investigations.} We assumed a metallicity of [Fe/H] = $-1.8$ dex as
representative of the RR Lyrae stars in all galaxies.

New distances were derived using the ($I$,$V-I$) PWR, with the
exception of DDO~210, for which only $B$ and $I$ catalogs are
available. A second estimate was obtained using the LMR, based on the
\citet{clementini03} calibration and updated as described in the
previous section. Table \ref{tab:table2} summarizes the results.
Column 2 presents the original literature values; we included only
estimates based on RR Lyrae stars, with the exception of Phoenix
\citep{hidalgo09,ordonez14}. Columns 3 and 4 present the same distance
estimates recalculated using the PWR and the updated LMR (Eq. 1). Both
methods show excellent agreement in most cases. Figure
\ref{fig:distnew} shows the differences between the literature values
and our updated values based on the PWR. Note that, overall, the
literature values are based on slightly higher mean metallicities than
adopted for our recalibration ([Fe/H] = $-1.5$ dex for Leo A, values
between [Fe/H] = $-1.6$ and $-1.8$ dex for the other galaxies). This
can account for up to few hundredths of a magnitude in distance
differences. Nevertheless, the change is small, at the level of at
most $\sim1\sigma$, where the error bars reflect realistic estimates
of systematic and random uncertainties associated with the method.

\section{Internally consistent distance determination beyond the nearest galaxies}
\label{sec:cluster}

Internally consistent determinations of the Hubble constant well
beyond the Local Group, based on cross-correlation of a variety of
independent techniques, are crucially important for establishing the
three-dimensional morphology of the Universe on the largest scales. By
association, such measurements provide indirect tests of the
consistency of the cosmological framework implied by SNe Ia
measurements. Distance measurements at these distances are
particularly important in the context of reducing the effects of
galaxies' peculiar velocities as well as regarding the inhomogeneity
of the matter distribution, which is the root cause of the difference
between the locally measured and the true, global values of the Hubble
constant \citep[e.g.,][]{turner92}.

SNe Ia are bright standard candles and, hence, in principle they
enable distance measurements out to cosmological scales ($z\sim
1$). However, to use SNe Ia as absolute distance indicators, their
absolute magnitudes must be calibrated using carefully selected
primary distance indicators such as Cepheid variables
\citep[e.g.,][]{riess16}. To check for the possible effects of unknown
systematics, those distances should be compared with a set of
independent distance measurements that do not rely on mutually
dependent calibrations of the distance ladder. Megamasers can be used
to directly measure distances to galaxies out to $\sim 50$~Mpc, which
is sufficiently distant to reduce the effects of peculiar velocities
\citep{reid09}. Here, we review several other independent techniques
that may potentially serve as important tests of the absolute distance
scale beyond the nearest galaxies.

\subsection{Clusters of galaxies}

Combined with X-ray data, observations of the Sunyaev--Zel'dovich
effect (SZE) affecting massive clusters of galaxies can be used to
infer the Hubble constant \citep{silk78}. This is because the X-ray
surface brightness depends on $S_X\propto \int n_e^2\Lambda {\rm
  d}\ell$, where $\Lambda$ is the X-ray cooling function determined
from the gas temperature $T_e$, and the SZE depends on $y \propto \int
n_e T_e {\rm d}\ell$. Assuming spherically shaped galaxy clusters,
line-of-sight distances can be expressed as ${\rm d}\ell \sim D_{\rm
  A} {\rm d}\theta$, where $D_{\rm A}$ is the angular diameter
distance and $\theta$ the cluster's angular size. By eliminating the
electron density, $n_e$, the angular diameter distance can be
expressed in terms of X-ray and SZ observables as $D_{\rm A} \propto
y^2/S_X$, assuming spherical symmetry. The effects of non-spherical
cluster morphologies can be reduced by averaging $H_0$ measurements
for many clusters.

\citet{reese02} determined distances to 18 clusters at $z=0.14-0.78$
to constrain the Hubble constant to $H_0=60^{+4}_{-3}({\rm
  statistical})^{+13}_{-18}({\rm systematic})$ km s$^{-1}$
Mpc$^{-1}$. \citet{bonamente06} obtained an updated result,
$H_0=76.9^{+3.9}_{-3.4}({\rm stat.})^{+10.0}_{-8.0}({\rm syst.})$ km
s$^{-1}$ Mpc$^{-1}$ using 38 clusters at $z=0.14-0.89$. These analyses
suggest that the dominant uncertainty affecting this method is already
determined by various systematic errors, such as inhomogeneities in
the intracluster medium \citep[e.g.,][]{kawahara08}. This may imply
that a better understanding of the physical state of the intracluster
medium \citep[e.g.,][]{hitomi16} is crucial to obtain more accurate
distance measurements based on this approach.

\subsection{Relative ages of old galaxies}

\citet{jimenez02} proposed to use relative ages of passively evolving
galaxies at a range of redshifts to directly constrain the Hubble
parameter, $H(z)$. Spectra of passively evolving galaxies are
dominated by main-sequence stars of $M\sim 1M_\odot$, for which the
relevant stellar evolution processes are well-understood. In addition,
relative ages can be determined more easily and with higher accuracy
than absolute ages, because several systematic effects are factored
out. Once the relative ages at different redshifts have been
determined, one can derive the Hubble parameter by employing the
relation $H(z)=-({\rm d}z/{\rm d}t)/(1+z)$. \citet{jimenez03} applied
this technique to Sloan Digital Sky Survey (SDSS) galaxies at $z<0.17$
to obtain $H_0=69\pm 12$ km s$^{-1}$ Mpc$^{-1}$, which is consistent
with other $H_0$ measurements. Further theoretical and observational
studies aimed at improving our understanding of galaxy spectra have
the potential to reduce the remaining uncertainties in the
current-best values of $H_0$.

\subsection{Gravitational waves as standard sirens}

Gravitational waves from inspiraling compact binaries are well placed
to become useful as absolute distance indicators
\citep{schutz86,holz05}. They are sometimes referred as `standard
sirens.' This is because masses of inspiraling and merging objects,
such as neutron stars and black holes, can be inferred from the shape
of the waveform (e.g., from the frequency and its time evolution),
which also determines the absolute strain amplitude (a dimensionless
parameter that depends on the strength of the tidal gravitational
field between both binary components). We can then directly measure
the luminosity distance to a gravitational-wave source, because the
observed strain is inversely proportional to the luminosity distance.

We can measure luminosity distances using gravitational-wave standard
sirens, but not the associated redshifts. Therefore, independent
observations of the objects' electromagnetic counterparts are usually
required to locate the host galaxy of a gravitational-wave event and
measure its redshift to constrain the Hubble constant. For instance,
\citet{dalal06} discussed the possibility of simultaneous observations
of short gamma-ray bursts and gravitational waves from merging
neutron-star binaries. They argued that several tens of such
observations can constrain the Hubble constant to the $\sim 2$\%
level.

A major advantage of this method is its `clean' physics. Assuming that
general relativity is valid, the waveforms of such merging events can
be computed easily from first principles. On the other hand, obtaining
observations of gravitational waves has been a significant challenge
for decades. Recently, the first observation of gravitational waves,
referred to as GW150914, from a pair of black holes has been reported
\citep{abbott16}. This has now finally opened up the possibility of
using gravitational waves as a tool to benchmark the cosmic distance
scale. Whereas it is not yet clear whether binary black hole mergers
produce any detectable electromagnetic counterparts for redshift
measurements, it may also be possible to use the spatial
cross-correlation of gravitational-wave sources and galaxies with
known redshifts to constrain the Hubble constant without any
observations of electromagnetic counterparts \citep{oguri16}. For this
cross-correlation technique to become viable, we need observations of
large numbers of gravitational waves over the entire sky, as well as
decent localizations of gravitational-wave sources, with an accuracy
of $\sim 1^\circ$. This will become possible with the next-generation
of gravitational-wave detectors.

\section{Gravitational lensing as a promising tool}
\label{sec:cosmodist}

Time delays of strongly lensed quasars provide a one-step measurement
of a cosmological distance, namely the time-delay distance to the lens
system \citep[e.g.,][]{Refsdal64, SuyuEtal10}. With additional
information from the stellar velocity dispersion of the lens, we can
further constrain the associated angular diameter distance
\citep{ParaficzHjorth09, JeeEtal15}. In a companion review, we have
described the recent cosmological constraints based on this approach,
particularly those based on the COSMOGRAIL \citep{CourbinEtal05,
  CourbinEtal11, EigenbrodEtal05, TewesEtal13b} and H0LiCOW
\citep{SluseEtal16, BonvinEtal16b, RusuEtal16b, SuyuEtal16,
  WongEtal16} projects. Here, we describe the future prospects for
this approach, focusing in particular on (i) expanding the time-delay
lens sample and (ii) follow-up observations, in light of upcoming
missions and surveys.

\subsection{Where are the time-delay lenses?}
\label{sec:cosmodist:lenssamp}

Quadruply lensed quasars (`quads') with ancillary data can each
provide a time-delay distance measurement of $\sim5-8$\%. However,
currently only a handful of quads are known at present. For example,
there are only four quads in the H0LiCOW sample of lenses for
cosmography \citep{SuyuEtal16}. There are several times more doubly
lensed quasars (`doubles'), but accurate and precise\footnote{Accuracy
  is a measure of how close the values are to the target or standard
  value; precision reflects their spread around the mean.} distance
measurements from these systems would be more challenging compared to
the equivalent determinations based on quads, since doubles have fewer
observational constraints compared with quads. For example, quad
systems can yield four image positions and flux measurements of
quasars as constraints, whereas double systems have half of that
number as constraints. In addition, quads can have three time-delay
measurements between the multiple images, which is substantially
better for cosmographic studies than the single time-delay measurement
available in doubles. Unless we can find more time-delay lenses,
especially quads, the statistical power of this approach will be
limited.

Fortunately, discoveries of hundreds if not thousands of time-delay
lenses are expected in current and future surveys and missions, with
$\sim13$\% of them being quads \citep{OguriMarshall10, LiaoEtal15}.
Below, we describe three ongoing imaging surveys that will likely
expand the current sample of quads by an order of magnitude:

\begin{enumerate}

\item {\bf Dark Energy Survey (DES):} The DES \citep{DES2016} uses the
  4 meter Mayall telescope at Cerro Tololo observatory in Chile. Its
  large-format CCD array of 74 detectors covers, in one single
  snapshot, an area of 2.2 degrees on a side
  \citep{FlaugherEtal2015}. After five years of operation the full
  survey will cover $\sim5000$ deg$^2$ in the $g, r, i, z$, and $Y$
  bands down to $r=24.3$ mag (10$\sigma$). DES has already discovered
  lensed quasars. Two were reported in \citet{AgnelloEtal15b} in the
  context of the SRIDES program \citep{AgnelloEtal2015c}. In addition,
  a doubly imaged quasar \citep{OstrovskiEtal2016} was found in the
  data taken during the first year of the survey. Many more are found
  at the time of writing this review, as the data obtained during the
  second year are analyzed. Given the area and depth of the survey,
  \citet{OguriMarshall10} predicted that a total of $\sim$1000 new
  lensed quasars should be found upon completion of the survey.

\item {\bf Hyper Suprime-Cam (HSC) Survey:} The HSC survey is a Subaru
  Strategic Program, using the newly installed HSC
  \citep{MiyazakiEtal12} on the Subaru 8.2 meter telescope, which is
  capable of imaging a large area of the sky in a single
  pointing\footnote{http://hsc.mtk.nao.ac.jp/ssp/}. The survey started
  in 2014 and is divided into three layers characterized by different
  areas and depths, i.e., wide, deep, and ultra-deep. In particular,
  we expect to find most of the new lenses in the wide survey, which
  will cover $\sim1400$ deg$^2$, mostly in equatorial regions, to
  $i$$\sim$26 mag in $grizy$ broad bands with excellent seeing
  ($\sim0.6''$ in the $i$ band). The first-year HSC survey data
  covering over 100 deg$^2$ in the five broad bands have recently been
  released to the public \citep{Aiharaetal17}. The expected number of
  lensed quasars in the HSC survey is $\sim 600$
  \citep{OguriMarshall10}.

\item {\bf Kilo-Degree Survey (KiDS):} The KiDS program is running at
  the European Southern Observatory's Paranal observatory in Chile
  \citep{Kuijken2015}. It uses the 2.6 meter VLT Survey Telescope
  (VST) in the $ugri$ optical bands and covers 10-degree wide strips,
  including an equatorial one between $10^{\rm h} 20^{\rm m} < {\rm
    RA} < 15^{\rm h} 50^{\rm m}$ and a southern strip going through
  the Galactic South Pole, $12^{\rm h} 00^{\rm m} < {\rm RA} < 03^{\rm
    h} 30^{\rm m}$. The total area covered is $\sim1500$ deg$^2$,
  i.e., similar to the HSC survey. The observing strategy is designed
  to ensure the best seeing in the $r$ band, which has a median seeing
  of $\sim 0.7''$. The depth in this band is $r=24.9$ mag (AB,
  5$\sigma$). Overall, KiDS should be as efficient as HSC in finding
  bright, strongly lensed quasars.

\end{enumerate}

Various new algorithms have been developed for finding galaxy-scale
lenses, many of which have been applied to these imaging surveys. Here
we focus on automated algorithms that are aimed at finding lensed
quasars, although there are also methods to find lensed galaxies
without quasars \citep[e.g.,][]{BoltonEtal06, GavazziEtal14,
  JosephEtal14, NapolitanoEtal15, ParaficzEtal16, ShuEtal16,
  Sonnenfeldetal17} or lens systems in general via visual inspection
by citizen scientists \citep[e.g.,][]{MarshallEtal16,
  MoreEtal16b}. While most of the first lensed quasars were found at
radio wavelengths through the Cosmic Lens All-Sky Survey
\citep{MyersEtal03}, the SDSS has yielded dozens of lensed quasars
through the SDSS Quasar Lens Search \citep[e.g.,][]{OguriEtal06,
  KayoEtal10, InadaEtal12, OguriEtal12} and, more recently, another 13
two-image lensed quasars \citep{MoreEtal16}. The advantage of using
SDSS is the availability of a spectroscopic quasar sample from which
one could search for lensed objects. Consequently, lenses obtained
from such a search will guaranteed be lensed quasars, rather than
lensed galaxies (without quasars), which are more numerous than lensed
quasars but not time-varying. Current imaging surveys such as HSC,
DES, and KiDS do not include a spectroscopic component as part of the
surveys. Therefore, recent searches necessarily rely on color
information by selecting objects with colors and morphologies
compatible with lensed active galactic nuclei
\citep[e.g.,][]{JacksonEtal12, AgnelloEtal15a, AgnelloEtal15b,
  OstrovskiEtal2016, SchechterEtal16}. In addition, as advocated by
\citet{MarshallEtal09}, for an object to qualify as a lens, it must
have a physical lens mass model. This is the basis of the model-based
search algorithms such as CHITAH \citep{ChanEtal15} and LensTractor
(Marshall et al., in prep.). Moreover, \citet{KochanekEtal06b} have
proposed to use quasar variability to find lens systems by looking for
clustered (possibly blended) variable sources corresponding to the
multiple quasar images. Our expectation is that these surveys will
each contain hundreds to about a thousand lensed quasars
\citep{OguriMarshall10}, and efforts are underway to find these.

Two surveys planned for execution the 2020s should yield yet another
order-of-magnitude increase in the number of lenses compared with
current surveys:

\begin{enumerate}

\item {\bf Euclid:} The first space-based, wide-field cosmological
  survey will be the ESA/NASA {\sl Euclid} 1.2 meter telescope
  \citep{LaureijsEtal11}. Once placed at the L2 Sun--Earth Lagrange
  point in 2020, this satellite will be equipped with a 0.5$\times$0.5
  deg$^2$ optical imager, which uses a single broad filter covering
  the $R$+$I$+$Z$ bands, as well as a near-infrared imager covering
  the same field of view to perform imaging in the $Y$, $J$, and $H$
  bands. In total, {\sl Euclid} will cover $\sim 15,000$ deg$^2$ of
  extragalactic sky down to a 10$\sigma$ optical AB magnitude of 24.5
  mag and down to a 5$\sigma$ AB magnitude of 24.0 mag in each of the
  three near-infrared bands. An additional deep survey will include 40
  deg$^2$ of sky two magnitudes deeper than the wide survey. {\sl
    Euclid} will image a total of 12 billion sources (3$\sigma$) and
  it will obtain near-infrared slitless spectra of 35 million of
  these. With its excellent image quality and point-spread-function
  (PSF) stability, {\sl Euclid} is optimized for weak-lensing
  tomography and it will also implementation of other cosmological
  probes such as baryon acoustic oscillations, the integrated
  Sachs--Wolfe effect, galaxy cluster counts, and redshift space
  distortions. With its sharp PSF ($0.18''$ FWHM), {\sl Euclid} will
  also be a superb machine to discover gravitationally lensed quasars
  with an image quality that, on its own, already allows one to
  constrain lens models.

\item {\bf Large Synoptic Survey Telescope
  (LSST)\footnote{https://www.lsst.org/}:} The LSST is an ambitious
  new survey telescope that is under construction in Chile. The
  telescope has a primary mirror diameter of 8.4 meters, with an
  effective aperture of 6.7 meters (because of the tertiary mirror
  area in the middle of the primary--tertiary mirror and some
  obscuration). It will rapidly survey the entire visible southern sky
  of $\sim 20,000$ deg$^2$ twice each week, for 10 years. Each patch
  of sky it will be visited 1000 times during the survey, in six
  filters, $ugrizy$. The survey will reach 24.5 mag depth in a single
  visit (30 seconds) and 27.5 mag co-added depth. The four primary
  science objectives of the LSST that drive the survey design are (i)
  constraining dark energy and dark matter, (ii) taking an inventory
  of the solar system, (iii) exploring the transient optical sky, and
  (iv) mapping the Milky Way. LSST will be revolutionary in the area
  of gravitational lensing. Thousands of lensed quasars will be found
  by LSST \citep{OguriMarshall10}, and at least hundreds of them will
  have well-defined time delays measurable directly from the survey
  data alone \citep{LiaoEtal15}.

\end{enumerate}

\subsection{Observational follow-up requirements and analysis}
\label{sec:cosmodist:followup}

\subsubsection{Confirmation}
\label{sec:cosmodist:followup:confirm}

Both high angular-resolution images and spectroscopy would help reveal
whether the lens candidates, found in ground-based imaging surveys,
are gravitational lenses or other astrophysical systems.

With high-resolution images, one can check whether the multiple image
features are consistent with lensing morphology. While {\sl Euclid}
images are of high-resolution by default, space-based or
adaptive-optics ground-based images could be used to confirm
candidates from ground-based imaging surveys. Nonetheless, some
systems observed with high-resolution imaging could still be difficult
to decipher, for example, a doubly-imaged quasar and a
star--quasar--galaxy chance alignment could look similar in
high-resolution images, in which case spectroscopy is needed for
confirmation.

Identical spectra of multiple components/features of an object would
confirm the object to be a strongly lensed system, with the multiple
components originating from the same source. Spectroscopic
confirmation could further yield spectroscopic redshift measurements
of the lens and/or the source, which are crucial pieces of information
to convert the angular quantities which we measure to physical
quantities.

\subsubsection{Follow-up observations}
\label{sec:cosmodist:followup:obs}

With ground-based and upcoming space-based wide-field surveys, not
only does finding lensed quasars become a fairly easy task, but also
some of the follow-up observations required to confirm (or not) the
candidates come for free. KiDS, DES, LSST, and all deep, multi-band
ground-based surveys have at least one of their bands optimized for
weak-lensing measurements. As a consequence, measuring the colors of
the individual quasar images can be done directly from the survey
data. In many cases, detecting the lensing galaxy or galaxies among
the quasar images is possible from the weak-lensing quality
images. However, additional ingredients are necessary to use the
confirmed lensed quasars for cosmological purposes.

\begin{enumerate}

\item {\bf The lens redshift and velocity dispersion:} In the era of
  modern wide-field surveys, the survey data on their own offer the
  required material to deblend the lensed images of the quasar and fit
  a crude lens model. More difficult are measurements of the lens
  redshifts, since they are often hidden in the glare of the quasar
  images, so that this requires long exposure times on 10 meter class
  telescopes. In COSMOGRAIL, \citet{EigenbrodEtal06, EigenbrodEtal07}
  measured 15 lens redshifts based on 2 hour exposure times and slit
  spectroscopy with the VLT. To separate the spectra of the quasar
  images from that of the lens galaxy, the data are spatially
  deconvolved \citep{CourbinEtal00} using the spatial information
  contained in the spectra of PSF stars, obtained in multi-slit
  mode. The same technique can be used to measure the velocity
  dispersion of the lensing galaxy. This requires, however, longer
  exposure times and an instrumental setup with higher spectral
  resolution. With the advent of integral-field spectrographs, in
  combination with adaptive optics, or mounted on the {\sl James Webb
    Space Telescope} ({\sl JWST}), it will become possible to map the
  full velocity fields and the velocity-dispersion fields of lens
  galaxies, hence providing a new wealth of observational constraints
  useful in breaking the degeneracies inherent to lens
  models. Figure~\ref{Fig:EC} illustrates what is possible already
  with the near-infrared two-dimensional SINFONI spectrograph mounted
  on the VLT.

\begin{figure}[t!]
\centering
\includegraphics[width=11.7cm]{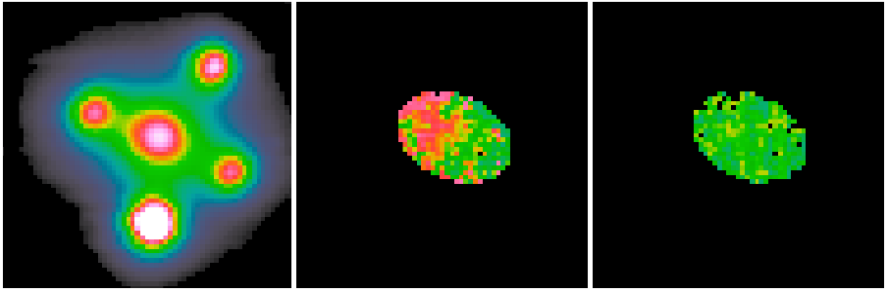}
\caption{(Left) Reconstructed image of the Einstein Cross, Q2237+080,
  seen with the near-infrared SINFONI integral-field spectrograph
  mounted on the VLT. The field of view is 3$''$ on a side. North is
  up and East to the left. No adaptive optics was used to obtain this
  6 hour integration with 0.5$''$ seeing. (Middle) Radial-velocity
  field showing a small rotation signal. The maximum velocity in this
  1$''$ field is 60 km s$^{-1}$. (Right) Line-of-sight
  velocity-dispersion field, which is relatively flat with a mean
  value of $\sigma_r = 160\pm 20$ km s$^{-1}$ (From Eigenbrod 2009,
  PhD Thesis).}
\label{Fig:EC}
\end{figure}

\item {\bf High-resolution imaging of the lensed quasar host:} Quasar
  time delays depend on the slope of the lens mass at the position of
  the quasar images, which is the quantity we need to constrain with
  observations. One route to such a mass-slope measurement is use of
  the Einstein ring formed by the quasar host galaxy. If the latter is
  radially extended, the leverage on the mass slope of the lens is
  sufficiently good to fit simple lens models such as power-laws or
  more physically justified models involving stellar and dark matter
  components \citep[][their Fig. 2]{SuyuEtal14}. Fortunately, such
  rings are visible in most lensed quasars with time delays, provided
  that deep and high-resolution images can be taken, as in the H0LiCOW
  project \cite[e.g.][]{SuyuEtal16}. Notably, the cosmological
  constraints derived using {\sl HST} imaging can be further improved
  with Keck adaptive optics \cite[][]{ChenEtal16}. {\sl JWST} and the
  new generation of ground-based 30--40 meter-class telescopes have
  the potential to revolutionize the field thanks to the levels of
  detail achievable in the images of Einstein rings
  \citep{MengEtal15}.

\item {\bf Line-of-sight effect:} While the multiple images of lensed
  quasars are formed by individual galaxies, the details of the image
  configuration and the precise slope of the total potential well at
  the image positions depend on the contributions from all massive
  objects up to the quasar redshift. Lens galaxies are often found in
  groups, i.e., they may share the dark matter halo of the group they
  belong to, in addition to their own halo. The contribution of the
  group to the total deflection field must be evaluated. If it is
  massive enough, deriving a mass map is possible from a weak-lensing
  analysis \cite[e.g.][]{KneibEtal00}. However, one must also account
  for the contribution of all other masses along the line of
  sight. Currently, this is done by comparing galaxy counts measured
  in the field of view of strongly lensed quasars
  \cite[e.g.][]{FassnachtEtal11} with galaxy counts in cosmological
  simulations \citep{HilbertEtal07,HilbertEtal09}. Lines of sight are
  chosen in the simulations with the same {\it relative} galaxy
  density as in the observations and used to derive a probability
  distribution function for the mass density toward a given lensed
  quasar \cite[][]{SuyuEtal10, GreeneEtal13, SuyuEtal13,
    RusuEtal16}. Other, more sophisticated methods
  \citep{CollettEtal13,McCully2016} have the potential to reduce the
  uncertainty along the line of sight to that dominated by statistical
  errors only. All the above methods require multiband imaging to
  measure photometric redshifts and stellar masses. Ideally,
  multi-object spectroscopy is desirable to infer in detail the mass
  contributions from all galaxies in the immediate vicinity of the
  lens ($< 15''$), as done recently for HE~0435-1223
  \citep{SluseEtal16}.

\item {\bf Time-delay measurements:} Large-scale sky surveys and
  discoveries of thousands of new lensed quasars potentially allow a
  drastic reduction in the random uncertainties affecting distance
  estimates using time delays. However, current time-delay
  measurements, e.g., based on COSMOGRAIL, need years of photometric
  monitoring. The reason for this is that the time-delay precision is
  limited by microlensing by stars in the lensing galaxy, which act as
  secondary lenses in each quasar image. As stars move along the line
  of sight to the quasar images, they affect the quasar images'
  apparent brightnesses. This results in a flickering of the lensed
  images on time-scales similar to those of the intrinsic quasar
  variations and with similar (or sometimes larger) amplitudes. Our
  ability to measure time delays therefore depends on our
  effectiveness in disentangling microlensing (different in each
  quasar image) from intrinsic quasar variations (identical in each
  quasar image). Obviously, it would be much easier to separate the
  microlensing variations from the intrinsic quasar variations if they
  were acting on different temporal time-scales. This is, in fact,
  possible! Quasars show variations on very short time-scales, much
  shorter than those typical of microlensing. The {\sl Kepler} light
  curves of active galactic nuclei reveal that essentially all quasars
  exhibit variations on time-scales of a week or two
  \citep{MushotzkyEtal11}. Since microlensing is present on scales of
  a few years, the microlensing versus quasar variations can be
  separated in frequency, provided that the low-amplitude and fast
  intrinsic quasar variations can be detected and measured. Typical
  slow quasar variations over time-scales of months have amplitudes on
  the order of 0.5 mag, whereas faster, week-long variations usually
  only span a few mmag (r.m.s.). For this reason, the future of lens
  monitoring may reside in high signal-to-noise and high-cadence
  monitoring, e.g., with daily observations of lenses with a 2--4
  meter-class telescope. Such observations allow one to catch the
  quasar variations, almost insensitive to long time-scale
  microlensing. Time-delay measurements become possible in 1 or 2
  years per object instead of 10--15 years, as is currently the case
  for smaller telescopes and a coarser temporal sampling of one or two
  epochs per week. However, given the signal-to-noise ratio needed
  (1000 per quasar image and per epoch), it will be difficult to
  monitor more than 20 lenses per night at any given telescope. In
  addition, measuring time delays longer than 90 days is not possible
  with this high-cadence mode over 1 or 2 years, since the time delay
  becomes comparable to the object's visibility period. Long-term
  monitoring with large telescopes capable of following hundreds of
  lenses per night, like the LSST, is therefore the other
  complementary route to enable large numbers of precise and accurate
  time-delay measurements.

\end{enumerate}

\subsubsection{Analysis and lens modeling}
\label{sec:cosmodist:followup:mod}

Modeling the lens-mass distribution and accounting for the mass
structure along the line of sight to the quasar source are crucial to
convert time delays into distance measurements. Deep imaging which
reveals the Einstein ring is important to precisely constrain the
lens-mass distribution near the location of multiple images, where it
matters for cosmography. While this has been done mostly in the
optical/infrared with thousands of intensity pixels
\citep[e.g.,][]{KochanekEtal01, KoopmansEtal03, DyeWarren05,
  SuyuEtal09, SuyuEtal13, FadelyEtal10, BirrerEtal16, WongEtal16},
interferometric radio data have also been employed
\citep[e.g.,][]{KochanekNarayan92, EllithorpeEtal96,
  WucknitzEtal04}. The recent completion of the Atacama Large
Millimeter/submillimeter Array (ALMA), which provides
high-sensitivity and high angular-resolution imaging, opens up a new
avenue for modeling the Einstein rings \citep[e.g.,][]{HezavehEtal13,
  BussmannEtal15, RybakEtal15, HezavehEtal16}.

To model the observed images, simultaneous determination of the source
surface brightness and the lens-mass distributions is needed.
Numerous modeling approaches have been developed over the past decade
\citep[e.g.,][]{WarrenDye03, SuyuEtal06, BarnabeKoopmans07,
  JulloEtal07, JulloKneib09, SuyuEtal09, VegettiKoopmans09,
  SuyuHalkola10, Oguri10, CollettAuger14, TagoreKeeton14,
  BirrerEtal15, NightingaleDye15, RybakEtal15, HezavehEtal16,
  TagoreJackson16}, some of which have been used recently to measure
distances to lensed quasars
\citep[e.g.,][]{SuyuEtal10,SuyuEtal13,SuyuEtal14, BirrerEtal16,
  WongEtal16}. In order to avoid confirmation bias, the cosmographic
analysis of the H0LiCOW collaboration \citep{SuyuEtal16} is blinded
\citep{SuyuEtal13,BonvinEtal16b,WongEtal16}. For cosmographic
inference, each lens takes an expert modeler weeks of analysis owing
to the high computational load caused by the large numbers of data
points (image pixels) and parameters (dozens) involved. While parallel
computing can reduce the computational time, frequent manual input
from the expert modeler is often required. Automating the full
modeling procedure is challenging given the complexity of lens
modeling, but it is worth pursuing in the era of hundreds/thousands of
lenses.

To break lens-modeling degeneracies caused by the mass-sheet
degeneracy and source-position transformation \citep[][]{FalcoEtal85,
  GorensteinEtal88, SchneiderSluse13, SchneiderSluse14, XuEtal16},
additional information such as the stellar kinematics of the lens
galaxy is needed. However, most of the lens modeling software suites
tend to make use of only the lensing information and incorporate the
kinematic information separately. \citet{BarnabeKoopmans07} and
\citet{BarnabeEtal11} provide self-consistent modeling of
three-dimensional axisymmetric lens-mass distributions, although the
cosmological implications of such a modeling strategy are yet to be
explored. We expect that two-dimensional kinematic maps of lens
galaxies would provide complementary information to break
lens-parameter degeneracies, especially for two-image systems. Future
lens modeling would benefit from including such kinematic
information. As a demonstration, \citet{SuyuEtal14} found a $\sim 4$\%
discrepancy in the time-delay distance measurements between different
lens mass models of a strong lens system owing to lensing
degeneracies. Incorporating an aperture-averaged velocity dispersion
measurement partially breaks the lensing degeneracy and reduces the
discrepancy to $\sim 1$\%. In addition, kinematic data are the key
ingredient for measuring the angular diameter distance to the lens
\citep{JeeEtal15}. This distance measurement substantially improves
the cosmographic constraints from time-delay distances alone, making
time-delay lenses an even more competitive cosmographic probe
\citep{JeeEtal16}.

In terms of quantifying line-of-sight structures, current approaches
for cosmography are based on relative galaxy counts around a strong
lens compared with those of simulations. With sufficient photometric
and spectroscopic information of the structures along the line of
sight, one could (in principle) attempt a full light-cone
reconstruction of the mass. Steps into this direction have been
carried out by, e.g., \citet{CollettEtal13}, \citet{McCullyEtal14},
and \citet{McCully2016}. This has the potential of achieving more
precise and accurate inferences about the external convergence.

\section{Future outlook}

The detailed narrative provided in this review should have impressed
upon the reader that efforts to establish an internally consistent
distance framework from the nearest galaxies to the highest redshifts
are indeed coming together on all scales. Analysis, theoretical, and
observational developments are indeed very promising, with significant
improvements in both accuracy and precision anticipated in the next
few years.

We are at the dawn of a new era for variability studies. Current and
future surveys are about to yield an unprecedented amount of
information on the variable stars populating the Local Group. The
European Space Agency's {\sl Gaia} mission will result in the
discovery of thousands of new RR Lyrae stars in the Milky Way halo,
and it will nail down the thorny systematics affecting the calibration
of the luminosity--metallicity relation.

Going to near-infrared and longer wavelengths may also enable us to
reduce the uncertainties in the distances to Local Group galaxies. At
present, 2--3\% distance accuracy is already achievable for distance
estimates to the LMC, which may soon be improved to $\sim 1$\%. For
instance, the Carnegie Hubble Program, using data from the warm {\sl
  Spitzer} mission, derived $(m-M)_0^{\rm LMC} = 18.477 \pm 0.034$ mag
\citep{free12}. \citet{ripe12} used near-infrared {\sl VISTA}
observations to derive $(m-M)_0^{\rm LMC} = 18.46 \pm 0.03$ mag, while
\citet{inno13} found $(m-M)_0^{\rm LMC} = 18.45 \pm 0.02$ (stat.) $\pm
0.10$ (syst.) mag based on optical/near-infrared PLR analysis of a
large sample of fundamental-mode LMC Cepheids. These distances are
within the mutual uncertainties of the direct, geometric distance
determination based on eclipsing binaries obtained by
\citet{pietrzynski13}. The most recent SMC distance determination,
which is based on mid-infrared analysis of fundamental-mode Cepheids,
yields $(m-M)_0^{\rm mid-IR} = 18.96 \pm 0.01 \mbox{ (stat.)} \pm
0.03$ (syst.) mag \citep[][see also \citealt{grac14}]{scow16}. Extant
and upcoming data will soon allow a systematic characterization of
variable stars, thus reducing error bars and fixing the first rungs of
the distance ladder.

Water maser measurements, first applied to obtain a geometric distance
to NGC 4258 \citep[][see also Table
  \ref{recommendations.tab}]{herr99}, have since been extended to
other nearby systems. Preliminary efforts to determine a
water-maser-based distance to M33 have thus far resulted in $D_{\rm
  M33} = 750 \pm 140 \pm 50$ kpc---$(m-M)_0^{\rm M33} =
24.38^{+0.49}_{-0.64}$ mag (total uncertainty)---where the first
uncertainty in the linear distance determination is related to
uncertainties in the H{\sc i} rotation model adopted and the second
uncertainty comes from proper-motion measurements
\citep[cf.][]{degr13}.

The LSST will produce a flood of data that will allow us to discover a
close-to-complete sample of RR Lyrae stars out to hundreds of kpc,
mapping the Milky Way's halo and its substructures. Moreover, the next
generation of extremely large telescopes will push the discovery of RR
Lyrae stars to few Mpc, and of classical Cepheids out to $\sim$100
Mpc.

Very Long Baseline Interferometry will enable geometric distance
determination out to more than 100 Mpc, including to NGC 4258, M33,
UGC 3789, and NGC 6264. Combined with {\it a priori} information on a
galaxy's rotation curve and its inclination with respect to our line
of sight, we can construct a slightly warped `tilted-ring' model of
the galaxy's dynamical structure. This, in turn, allows correlation of
the angular proper motion measurements with the rotational velocity
information obtained in linear units and, thus, provides an
independent distance measurement.

Simultaneously, the Megamaser Cosmology Project
\citep[e.g.,][]{reid09,reid13,braa10} aims at using extragalactic
maser sources to directly measure $H_0$ in the Hubble flow, which is
clearly very challenging at distances beyond 100 Mpc. The project
team's preliminary results look promising, however: using NGC 6264 ($D
= 144 \pm 19$ Mpc) as a benchmark, they find $H_0 = 68 \pm 9$ km
s$^{-1}$ Mpc$^{-1}$ \citep{kuo13}, which is indeed very close to the
current best determinations of $H_0$ based on a variety of independent
measures. This looks like a promising way forward to eventually
construct an internally consistent distance ladder with well-defined
uncertainties out to the Hubble flow \citep[see also][]{thevenin17}.

With the newly discovered lenses expected from upcoming surveys, the
anticipated observational follow-up, and analysis developments, we
expect that measuring distances to a sample of $\sim50$ lenses should
be feasible within the next few years. \citet{JeeEtal16} showed that
with a sample of 55 bright lenses from current and future imaging
surveys, the resulting cosmological constraints will be highly
complementary and competitive to other cosmological probes, including
the cosmic microwave background, baryon acoustic oscillations, and SNe.

\begin{acknowledgement}
This research was partially supported by the National Natural Science
Foundation of China (NSFC; grants U1631102, 11373010, and 11633005 to
R.d.G.), the Spanish Ministry of Economy and Competitiveness (MINECO;
grant AYA2014-56795-P to M.M. and C.E.M.-V.), the Max Planck Society
through the Max Planck Research Group (S.H.S.), and the Swiss National
Science Foundation (SNSF; F.C.). We thank ISSI-BJ for hospitality and
an engaging workshop. We also acknowledge both referees for their
constructive reviews.
\end{acknowledgement}

\bibliographystyle{spbasic}
\bibliographystyle{spmpsci}
\bibliographystyle{spphys}

\end{document}